%% file: main.tex
\newif\iftiming\timingfalse
\newif\ifmytheorem\mytheoremfalse
\newif\ifelsa\elsafalse
\newif\iflncs\lncstrue
\newif\ifanswered\answeredfalse
\newif\iflong\longfalse
\newif\ifworking\workingfalse
\newif\ifprivate\privatefalse
\newif\ifslides\slidesfalse
\newif\ifreport\reportfalse  
\newif\ifok\oktrue
\newif\ifexplicit\explicitfalse
\newif\ifsigplan\sigplanfalse
\newif\ifold\oldfalse
\newif\ifproofs\proofsfalse
\newif\iftrace\tracefalse

\def\VERSION{}

\documentclass[envcountsame,orivec]{llncs}[2017/09/04]
\pdfoutput=1

\input{tikz.diagram}
\input{preamble}
\input{macros}

\input{macros-local}

\title{Ready, set, Go!}
\subtitle{Data-race detection and the Go language\thanks{Supported by the bilateral project UTF-2018-CAPES-Diku/10001 ``Modern Refactoring''.}}
\author{Daniel Schnetzer Fava~~~~ \qquad\qquad\quad Martin Steffen~~~~~~\\ 
\href{mailto:danielsf@ifi.uio.no}{\texttt{danielsf@ifi.uio.no}} \qquad\quad
\href{mailto:msteffen@ifi.uio.no}{\texttt{msteffen@ifi.uio.no}}}

\institute{Dept. of Informatics, University of Oslo}

\def\WHERE{main}
\begin{document}

\maketitle{}

\input{abstract}
\input{intro}
\input{background}
\input{race}
\input{race-gc}
\input{comparison}
\input{discussion}

\input{related}
\input{conclusion}

\bibliographystyle{apalike}
{\small
\bibliography{extracted}
}

\newpage

\appendix
\def\WHERE{appendix}
{\def\VERSION{.naked}
  \input{strongsemantics}
}
\input{append}

\ifworking
\input{notes}
\listoffixmes
\fi

\printindex
\end{document}


%% file: tikz.diagram.tex
\newcommand{\myhb}{}
\newcommand\dop[6][]{
  \draw [fill] (#4+0.05,#5) arc (0:180:0.05) -- cycle;
  \draw [fill] (#4-0.05,#5-0.3) arc (180:360:0.05) -- cycle;
  \draw [fill] (#4-0.05,#5) rectangle node (#2s) {} ++(0.1, 0.001);
  \draw [fill] (#4-0.05,#5-0.3) rectangle node (#2e) {} ++(0.1, 0.001);
  \draw [] (#2s.center) -- (#2e.center)
  node [#6=2mm, midway] (1) {#3};
}
\newcommand\pop[6][]{
  \draw [fill] (#4,#5) circle [radius=0.05] node (#2) {}
  node [#6=2.5mm] (1) {#3};
}

%% file: preamble.tex
\makeatletter
\def\NAT@spacechar{~}
\makeatother

\usepackage[title]{appendix}
\usepackage{subfig}[2002/07/30]  
\usepackage{soul}
\usepackage{etex}
\ifslides\else
\usepackage[dvipsnames]{xcolor}
\fi

\usepackage[author=]{fixme}[2013/01/28]

\usepackage[lighttt]{lmodern} 

\ifworking
\usepackage[numbers]{natbib}
\else
\usepackage[numbers]{natbib}
\fi

\usepackage{dsfont}

\fxusetheme{color}

\FXRegisterAuthor{ms}{ems}{ms}
\FXRegisterAuthor{df}{edf}{df}

\fxsetup{mode=multiuser}
\fxsetup{targetlayout=color}
\fxsetface{margin}{\tiny}

\fxsetup{status=draft}

\fxuseenvlayout{color}
\fxsetup{envface=\scriptsize\tt}  

\fxsetup{inlineface=\bfseries}
\fxsetup{envlayout=color}

\usepackage{shuffle}    

\usepackage{tikz}
\usetikzlibrary{automata}
\usetikzlibrary{cd}
\usetikzlibrary{calc}

\usetikzlibrary{matrix,positioning,tikzmark}
\usepackage{tikz-qtree}                   
\usetikzlibrary{matrix,arrows}  
\usetikzlibrary{chains,fit,shapes}
\usetikzlibrary{arrows,automata}
\usetikzlibrary{graphs}

\usetikzlibrary{arrows, decorations.markings}

\tikzstyle{inititem}=[fill=blue!10]
\tikzstyle{completeitem}=[fill=red!10]
\tikzstyle{conflictitem}=[fill=red!45]
\tikzstyle{initcompleteitem}=[fill=violet!20]

\tikzstyle{action}=[rectangle,draw,thick,rounded corners]
\tikzstyle{abefore}=[very thick,beamerblue,dotted]
\tikzstyle{po}=[very thick,beamerblue,dotted]
\tikzstyle{rf}=[very thick,beamerred,dotted]
\tikzstyle{dd}=[very thick,green,dotted]  

\usepackage{colortbl}
\definecolor{umbra}{rgb}{0.8,0.8,0.5}
\usepackage{comment}
\usepackage{hyperref}
\usepackage{rotating}
\ifslides\else
\usepackage[fleqn]{amsmath}
\fi
\usepackage{amssymb}
\usepackage{amstext}
\usepackage{cprotect}
\usepackage{mathptmx} 

\usepackage{mathtools}   
\usepackage{stmaryrd}
\usepackage{mathptmx} 
\DeclareMathAlphabet{\mathcal}{OMS}{cmsy}{m}{n}
\input{prooftree}
\usepackage{varioref}
\ifprivate\else

\fi
\usepackage{xifthen}

\usepackage{listings}[2002/04/01]  

\usepackage{lstlanggo}
\usepackage{lstpseudo}

\lstdefinestyle{mmgo.k}{
  basicstyle=\ttfamily\normalsize,
  morekeywords={rule},
  otherkeywords={=>},
  emph=[1]{G, goroutine, id, k, sigma, HB, S},
  emphstyle=[1]{\color{NavyBlue}},
  emph=[2]{W},
  emphstyle=[2]{\color{Dandelion}},
  emph=[3]{C, chan, ref, type, forward, backward},
  emphstyle=[3]{\color{ForestGreen}},
  emph=[4]{requires},
  emphstyle=[4]{\color{Mahogany}},
  escapeinside={(*@}{@*)},
  mathescape=false,
}
\usepackage{graphics,color}
\usepackage{graphicx}
\graphicspath{{./figures/}}    

\usepackage{makeidx}
\makeindex


%% file: prooftree.tex
\def\introrule{{\cal I}}\def\elimrule{{\cal E}}

\newdimen\proofrulebreadth \proofrulebreadth=.05em
\newdimen\proofdotseparation \proofdotseparation=1.25ex
\newdimen\proofrulebaseline \proofrulebaseline=2ex
\newcount\proofdotnumber \proofdotnumber=3
\let\then\relax
\def\hfi{\hskip0pt plus.0001fil}
\mathchardef\squigto="3A3B
\newif\ifinsideprooftree\insideprooftreefalse
\newif\ifonleftofproofrule\onleftofproofrulefalse
\newif\ifproofdots\proofdotsfalse
\newif\ifdoubleproof\doubleprooffalse
\let\wereinproofbit\relax
\newdimen\shortenproofleft
\newdimen\shortenproofright
\newdimen\proofbelowshift
\newbox\proofabove
\newbox\proofbelow
\newbox\proofrulename
\def\shiftproofbelow{\let\next\relax\afterassignment\setshiftproofbelow\dimen0 }
\def\shiftproofbelowneg{\def\next{\multiply\dimen0 by-1 }%
\afterassignment\setshiftproofbelow\dimen0 }
\def\setshiftproofbelow{\next\proofbelowshift=\dimen0 }
\def\setproofrulebreadth{\proofrulebreadth}

\def\prooftree{
\ifnum  \lastpenalty=1
\then   \unpenalty
\else   \onleftofproofrulefalse
\fi
\ifonleftofproofrule
\else   \ifinsideprooftree
        \then   \hskip.5em plus1fil
        \fi
\fi
\bgroup
\setbox\proofbelow=\hbox{}\setbox\proofrulename=\hbox{}%
\let\justifies\proofover\let\leadsto\proofoverdots\let\Justifies\proofoverdbl
\let\using\proofusing\let\[\prooftree
\ifinsideprooftree\let\]\endprooftree\fi
\proofdotsfalse\doubleprooffalse
\let\thickness\setproofrulebreadth
\let\shiftright\shiftproofbelow \let\shift\shiftproofbelow
\let\shiftleft\shiftproofbelowneg
\let\ifwasinsideprooftree\ifinsideprooftree
\insideprooftreetrue
\setbox\proofabove=\hbox\bgroup$\displaystyle 
\let\wereinproofbit\prooftree
\shortenproofleft=0pt \shortenproofright=0pt \proofbelowshift=0pt
\onleftofproofruletrue\penalty1
}

\def\eproofbit{
\ifx    \wereinproofbit\prooftree
\then   \ifcase \lastpenalty
        \then   \shortenproofright=0pt  
        \or     \unpenalty\hfil         
        \or     \unpenalty\unskip       
        \else   \shortenproofright=0pt  
        \fi
\fi
\global\dimen0=\shortenproofleft
\global\dimen1=\shortenproofright
\global\dimen2=\proofrulebreadth
\global\dimen3=\proofbelowshift
\global\dimen4=\proofdotseparation
\global\count255=\proofdotnumber
$\egroup  
\shortenproofleft=\dimen0
\shortenproofright=\dimen1
\proofrulebreadth=\dimen2
\proofbelowshift=\dimen3
\proofdotseparation=\dimen4
\proofdotnumber=\count255
}

\def\proofover{
\eproofbit 
\setbox\proofbelow=\hbox\bgroup 
\let\wereinproofbit\proofover
$\displaystyle
}%
\def\proofoverdbl{
\eproofbit 
\doubleprooftrue
\setbox\proofbelow=\hbox\bgroup 
\let\wereinproofbit\proofoverdbl
$\displaystyle
}%
\def\proofoverdots{
\eproofbit 
\proofdotstrue
\setbox\proofbelow=\hbox\bgroup 
\let\wereinproofbit\proofoverdots
$\displaystyle
}%
\def\proofusing{
\eproofbit 
\setbox\proofrulename=\hbox\bgroup 
\let\wereinproofbit\proofusing
\kern0.3em$
}

\def\endprooftree{
\eproofbit 
  \dimen5 =0pt
\dimen0=\wd\proofabove \advance\dimen0-\shortenproofleft
\advance\dimen0-\shortenproofright
\dimen1=.5\dimen0 \advance\dimen1-.5\wd\proofbelow
\dimen4=\dimen1
\advance\dimen1\proofbelowshift \advance\dimen4-\proofbelowshift
\ifdim  \dimen1<0pt
\then   \advance\shortenproofleft\dimen1
        \advance\dimen0-\dimen1
        \dimen1=0pt
        \ifdim  \shortenproofleft<0pt
        \then   \setbox\proofabove=\hbox{%
                        \kern-\shortenproofleft\unhbox\proofabove}%
                \shortenproofleft=0pt
        \fi
\fi
\ifdim  \dimen4<0pt
\then   \advance\shortenproofright\dimen4
        \advance\dimen0-\dimen4
        \dimen4=0pt
\fi
\ifdim  \shortenproofright<\wd\proofrulename
\then   \shortenproofright=\wd\proofrulename
\fi
\dimen2=\shortenproofleft \advance\dimen2 by\dimen1
\dimen3=\shortenproofright\advance\dimen3 by\dimen4
\ifproofdots
\then
        \dimen6=\shortenproofleft \advance\dimen6 .5\dimen0
        \setbox1=\vbox to\proofdotseparation{\vss\hbox{$\cdot$}\vss}%
        \setbox0=\hbox{%
                \advance\dimen6-.5\wd1
                \kern\dimen6
                $\vcenter to\proofdotnumber\proofdotseparation
                        {\leaders\box1\vfill}$%
                \unhbox\proofrulename}%
\else   \dimen6=\fontdimen22\the\textfont2 
        \dimen7=\dimen6
        \advance\dimen6by.5\proofrulebreadth
        \advance\dimen7by-.5\proofrulebreadth
        \setbox0=\hbox{%
                \kern\shortenproofleft
                \ifdoubleproof
                \then   \hbox to\dimen0{%
                        $\mathsurround0pt\mathord=\mkern-6mu%
                        \cleaders\hbox{$\mkern-2mu=\mkern-2mu$}\hfill
                        \mkern-6mu\mathord=$}%
                \else   \vrule height\dimen6 depth-\dimen7 width\dimen0
                \fi
                \unhbox\proofrulename}%
        \ht0=\dimen6 \dp0=-\dimen7
\fi
\let\doll\relax
\ifwasinsideprooftree
\then   \let\VBOX\vbox
\else   \ifmmode\else$\let\doll=$\fi
        \let\VBOX\vcenter
\fi
\VBOX   {\baselineskip\proofrulebaseline \lineskip.2ex
        \expandafter\lineskiplimit\ifproofdots0ex\else-0.6ex\fi
        \hbox   spread\dimen5   {\hfi\unhbox\proofabove\hfi}%
        \hbox{\box0}%
        \hbox   {\kern\dimen2 \box\proofbelow}}\doll%
\global\dimen2=\dimen2
\global\dimen3=\dimen3
\egroup 
\ifonleftofproofrule
\then   \shortenproofleft=\dimen2
\fi
\shortenproofright=\dimen3
\onleftofproofrulefalse
\ifinsideprooftree
\then   \hskip.5em plus 1fil \penalty2
\fi
}



%% file: macros.tex
\newcommand{\hbarrow}             {\rightarrow_{\mathsf{hb}}}   

\def\Cplusplus{{C\nolinebreak[4]\hspace{-.05em}\raisebox{.4ex}{\tiny\bf ++}}} 
\newcommand{\Cpluspluseleven}     {\Cplusplus 11}

\newcommand{\CCpluspluseleven}    {C/\Cpluspluseleven}

\newcommand{\Go}                  {Go}

\newcounter{counttodo}
\newcommand{\finishlater}       [1] {%
  \typeout{      >>>>>    Finish this later    <<<<< }}
\newcommand{\cleanup}[1]{\typeout{      >>>>> unify, clean up convention  <<<<< }}

\definecolor{beamergreen}{rgb}{0.0,0.40,0}    
\definecolor{beamerblue}{rgb}{0.2,0.2,0.7}  
\definecolor{beamerred}{rgb}{0.7,0.2,0.2}   
\definecolor{darkseagreen}{rgb}{0.56, 0.74, 0.56}

\newcommand{\myblue}[1]                 {\textcolor{beamerblue}{#1}}
\newcommand{\myred}[1]                  {\textcolor{beamerred}{{#1}}}

\newcommand{\gcloses}                  {\kw{close}}

\newcommand{\gisclosed}                {\mathit{closed}}
\newcommand{\eot}                      {\bot}                     

\newcommand{\gsend}[2]           {#1 \leftarrow #2}               
\newcommand{\greceive}[1]        {   \mathop{\leftarrow #1}}      

\newcommand{\gsendguard}[2]      {\gsend{#1}{#2} }
\newcommand{\greceiveguard}[2]   {\greceive{#1} }

\newcommand{\ngoroutine}[3]      {#1\angles{#2,#3}}      
\newcommand{\nslgoroutine}[2]    {#1\angles{#2}}         
\newcommand{\gchan}[2]           {#1[#2]}                
\newcommand{\gchanf}[2]          {#1_f[#2]}              
\newcommand{\gchanb}[2]          {#1_b[#2]}              

\newcommand{\lstate}               {\sigma}   
\newcommand{\lstateempty}        {{\lstate}_\bot}

\ifmytheorem
  \usepackage{ntheorem}
  \newtheorem{theorem}                      {Theorem}[section]
  \newtheorem{remark}[theorem]{Remark}

  \newtheorem{definition}      [theorem]    {Definition}

  \newtheorem{notation}        [theorem]    {Notation}

  \def\squareforqed{\hbox{\rlap{$\sqcap$}$\sqcup$}}
  \def\qed{\ifmmode\squareforqed\else{\unskip\nobreak\hfil
  \penalty50\hskip1em\null\nobreak\hfil\squareforqed
  \parfillskip=0pt\finalhyphendemerits=0\endgraf}\fi}
\fi

\makeatother

\newcommand{\casefont}            {\it}
\newcounter{casecounter}
\renewcommand{\thecasecounter}{\arabic{casecounter}}

\newcommand{\Casestring}          {Case}

\newcommand{\Subcasestring}       {Subcase}
\newcommand{\Subsubcasestring }   {Subsubcase}
\newcommand{\Subsubsubcasestring} {Subsubsubcase}
\newcommand{\Basecasestring}      {Base step}
\newcommand{\Inductioncasestring} {Induction step}

\newenvironment{namedcase}[1]{
  \smallbreak \pagebreak[3]
  \noindent #1 \\}%

\newenvironment{ruleset}{
  \noindent\mbox{}\noindent\hrulefill
  \begin{displaymath}
    \begin{array}[b]{l}}{
  \end{array}\end{displaymath}\hrulefill\mbox{}}

\newcommand{\ruleskip}   {\\ \\[-1.1em]}

\newcommand{\treefont}{\small}
\newcommand{\leaf}[1]{{\begin{array}{c} #1 \end{array}}}

\newcommand{\namedruletree}[3]{{
     \treefont
     \prooftree 
       {\leaf{#2}}
     \justifies
       #3
     \using \rn{#1}
     \endprooftree
     }}

\newcommand{\andalso}{\quad\quad}

\newcommand{\infrule}[3]{\namedruletree{#1}{#2}{#3}\andalso}

\newcommand{\infax}[2]{ 
  \treefont
  #2\quad\quad \rn{#1}\andalso
  \andalso
  }

\renewcommand{\infax}[2]{
  {\mbox{\treefont \ensuremath{#2}}}\quad\treefont\rn{#1}\andalso\andalso}

\newcommand{\rn}[1]{\mbox{\textsc{#1}}}        

\newcommand{\union}                            {\cup}

\newcommand{\nuin}[2]                           {\mathbf{\nu} #1\ #2}

\newcommand{\store}[2]           {#1:=#2}

\newcommand{\load}[1]            {\kw{load} #1}

\newcommand{\idshb}              {E_{\mathit{hb}}}

\newcommand{\kw}[1]                              {\mathrel{\mathtt{#1}}} 
\newcommand{\T}                                  {T}                     

\newcommand{\emptyconf}                          {\bullet}

\newcommand{\selects}                            {\sum}
\newcommand{\defaults}                           {\kw{default}}

\newcommand{\gos}                                {\kw{go}}

\newcommand{\lets}                               {\kw{let}}

\newcommand{\ins}                                {\kw{in}}
\newcommand{\ifs}                                {\kw{if}}          
\newcommand{\thens}                              {\kw{then}}       
\newcommand{\trues}                              {\kw{true}}
\newcommand{\falses}                             {\kw{false}}
\newcommand{\elses}                              {\kw{else}}        
\newcommand{\stops}                              {\kw{stop}}

\newcommand{\makes}                              {\kw{make}}
\newcommand{\makechans}[2]                       {\kw{make}(\kw{chan} #1,#2)}

\newcommand{\esend}[3]                           {\mygreen{#1(#2,#3)}}                       
\newcommand{\esendack}[2]                        {#1(#2)}                                    
\newcommand{\ewritea}[2]                         {\myred{(#1, #2)}}                          
\newcommand{\ereada}[1]                         {#1}                                        
 
\newcommand{\ewritesl}[3]                        {#1\bananas{#2{:=}#3}}                      

\newcommand{\trans}[1]                          {\xrightarrow{#1}}

\newcommand{\widetrans}[1]                      {~~\trans{#1}~~}

\newcommand{\transloc}[1]                       {\xrightarrow{\scriptscriptstyle#1}} 
\renewcommand{\transloc}[1]                     {\mathrel{\rightsquigarrow}}           

\newcommand{\eqdef}                             {\triangleq}
\newif\ifmyundef\myundeftrue
\ifsigplan
  \myundeffalse
\fi
\ifslides
  \myundeffalse
\fi
\ifmyundef
   
\fi

\newcommand{\sizeof}[1]                         {|\,#1\,|}
\newcommand{\len}[1]                            {|\,{#1}\,|}

\newcommand{\angles}[1]                         {\langle #1 \rangle}     

\newcommand{\lbanana}         {(\!\!|}
\newcommand{\rbanana}         {|\!\!)}
\newcommand{\llense}          {[\!\!\!\!\;(}
\newcommand{\rlense}          {)\!\!\!\!\;]}
\newcommand{\bananas}[1]      {\lbanana #1 \rbanana}
\newcommand{\lenses}[1]       {\llense #1 \rlense}

\newcommand{\fv}[1]                {\mathit{fv}(#1)}            

\newcommand{\suchthat}             {\mathrel{|}}
\newcommand{\rts}[1]               {\underline{#1}}             

\newcommand{\bnfdef}               {::=}
\newcommand{\bnfbar}               {\ \mathrel{|}\ }

\newcommand{\substfor}[2]             {[{#1}/{#2}]}            

\newcommand{\projectto}[1]         {\downarrow_{#1}}

\newcommand{\congstruct}           {\equiv}

\newcommand{\semsubsuper}[3]       {[\![#3]\!]_{#1}^{#2}}    

\newcommand{\p}                 {\mathrel{\vdash}}

\newcommand{\fn}                    {\mathit{fn}}                    
\newcommand{\fresh}                 {\mathit{fresh}}

\newcommand{\unit}                   {()}

\newcommand{\Texc}                   {\mathsf{E}}        

%% file: macros-local.tex
\newcommand{\nthreads}                      {\tau}
\newcommand{\nvars}                         {\nu}

\makeatletter
\newcommand{\pushleft}[1]{\ifmeasuring@#1\else\omit$\displaystyle#1$\hfill\fi\ignorespaces}
\makeatother

\newcommand{\TexcArg}[1]                    {\Texc\left({#1}\right)}
\newcommand{\incs}                          {inc}
\newcommand{\maxs}                          {max}

\newcommand{\vc}[1]                         {\mathds{#1}}
\newcommand{\vcleq}                         {\sqsubseteq}
\newcommand{\vcjoin}                        {\sqcup}
\newcommand{\vcbot}[1]                      {\bot_{#1}}
\newcommand{\vcinc}[1]                      {\incs_{#1}}
\newcommand{\epoch}[2]                      {{#1}@{#2}}

\newcommand{\hbat}[1]                       {\idshb@{#1}}

\newcommand{\idshbz}[1]                     {\idshb^{#1}}
\newcommand{\idshbzp}[1]                    {\idshb'^{#1}}
\newcommand{\ezwritesl}[3]                  {\bananas{#3,\, #1{:=}#2}}

\newcommand{\ind}                           {\bowtie}
\newcommand{\idxs}                          {\kw{idx}}
\newcommand{\idx}[2]                        {\idxs({#1},{#2})}

\newcommand{\tracepreo}                     {\sqsubseteq} 

\newcommand{\isSds}                         {\kw{isSd}}
\newcommand{\isRvs}                         {\kw{isRv}}
\newcommand{\funcIsSd}[2]{
  \ifthenelse{\isempty{#2}}%
    {{\isSds{#1}}}%
    {{\isSds{#1}~{#2}}}%
}
\newcommand{\funcIsRv}[2]{
  \ifthenelse{\isempty{#2}}%
    {{\isRvs{#1}}}%
    {{\isRvs{#1}~{#2}}}%
}

\newcommand{\wellform}[1]{%
  \ifthenelse{\isempty{#1}}%
    {{\vdash}}%
    {{\vdash{#1}}}%
}

\newcommand{\sds}                               {\kw{sd}}
\newcommand{\rvs}                               {\kw{rv}}

\newcommand{\sdop}[3]{%
  \ifthenelse{\isempty{#3}}%
    {\sds^{#1}\ {#2}}%
    {\sds^{#1}\ {#2}\ {#3}}%
}
\newcommand{\rvop}[3]{%
  \ifthenelse{\isempty{#3}}%
    {\rvs^{#1}\ {#2}}%
    {\rvs^{#1}\ {#2}\ {#3}}%
}

\newcommand{\makeop}[2]{%
  \ifthenelse{\isempty{#2}}%
    {\makes {#1}}%
    {\makes {#1}\ {#2}}%
}

\newcommand{\wrop}[2]                           {{#1}!{#2}}
\newcommand{\rdop}[2]                           {{#1}?{#2}}
\newcommand{\wrev}[3]                           {\myred{\event{#1}{\wrop{#2}{#3}}}}
\newcommand{\rdev}[3]                           {\myblue{\event{#1}{\rdop{#2}{#3}}}}

\renewcommand{\ewritea}[2]                      {\myred{(#1, !#2)}}    
\renewcommand{\ereada}[2]                       {\myblue{(#1,?#2)}}    

\newcommand{\event}[2]                          {({#2})_{#1}}

\newcommand{\sdev}[4]                           {\event{#1}{\sdop{#2}{#3}{#4}}}
\newcommand{\rvev}[4]                           {\event{#1}{\rvop{#2}{#3}{#4}}}

\newcommand{\acqs}                               {\kw{acq}}
\newcommand{\rels}                               {\kw{rel}}
\newcommand{\acq}[1]                             {\acqs(#1)}
\newcommand{\rel}[1]                             {\rels(#1)}
\newcommand{\tlock}[2]                           {#1[#2]}

\newcommand{\djit}                               {\textsc{Djit}$^+$}
\newcommand{\djitp}                              {\textsc{Djit}$\overset{+}{\text{.}\hspace*{0.3em}}$}

\newcommand{\fasttrack}                          {\textsc{FastTrack}}

\newcommand{\racerd}                             {\textsc{RacerD}}
\newcommand{\grace}                              {\textsc{Grace}}

\newcommand{\erwritesl}[4]                        {#1\bananas{#2{:=}#3}#4}

\newcommand{\phb}[1]                  {\p_{\mathsf{hb}}\ #1}

\renewcommand{\lstate}                {\idshb}
\newcommand{\idshbr}                  {\idshb^{r}}
\newcommand{\idshbrp}                 {\idshb'^{r}}
\renewcommand{\erwritesl}[4]          {#1\bananas{#4,\, #2{:=}#3}}

\ifelsa
\newcommand{\proofofref}{}
\newproof{zproofof}{Proof of \proofofref}
\newenvironment{proofof}[1]
 {\renewcommand{\proofofref}{#1}\zproofof}
 {\endzproofof}
\else

\fi

\tikzstyle{ppoedge}       = [->,very thick,MidnightBlue]
\tikzstyle{poedge}        = [->,very thick,black]
\tikzstyle{hbedge}        = [->,very thick,black]
\tikzstyle{rfedge}        = [->,very thick,red]
\tikzstyle{lwfedge}       = [->,very thick,OliveGreen]
\tikzstyle{ffedge}        = [->,very thick,OliveGreen]
\tikzstyle{fredge}        = [->,very thick,orange]

\tikzstyle{litmusdiagram} = [->,>=stealth',shorten >=1pt,auto,node distance = 2.1cm,auto]

\tikzset{
    ltevent/.style={
           rectangle,
           fill=none,
           rounded corners,
           draw=none,
           minimum height=2em,
           inner sep=8pt,
           text centered,
           },
}

\tikzset{
    ltthread/.style={
           rectangle,
           fill=none,
           rounded corners,
           draw=none,
           minimum height=2em,
           inner sep=8pt,
           text centered,
           },
}

\newcommand{\semobsl}[3]          {\semsubsuper{}{}{}}

\newcommand{\ereadfrom}[2]       {#1\lenses{#2}}

\tikzset{cddot/.style={circle,draw,fill,minimum size=0.5mm,inner sep=0.0mm}}
\tikzset{cdiagram/.style={node distance=1.5cm,shorten <= 1mm, shorten >= 1mm}}

%% file: abstract.tex
\begin{abstract}
Data races are often discussed in the context of lock acquisition and release,
with race-detection algorithms routinely relying on \emph{vector clocks} as a means of capturing the relative ordering of events from different threads.
In this paper, we present a data-race detector for a language with channel communication as its sole synchronization primitive,
and provide a semantics directly tied to the \emph{happens-before} relation, thus forging the notion of vector clocks.
\end{abstract}


%% file: intro.tex
\section{Introduction}
\label{sec:race.intro}

One way of dealing with complexity is by partitioning a system into
cooperating subcomponents.  When these subcomponents compete for resources,
coordination becomes a prominent goal.  One common programming paradigm is
to have threads cooperating around a pool of shared memory.  In this case,
coordination involves, for example, avoiding conflicting accesses to
memory.  Two concurrent accesses constitute a \emph{data race} if they
reference the same memory location and at least one of the accesses is a
write.  Because data races can lead to counter intuitive behavior, it is
important to detect them.

\index{data race} The problem of data-race detection in shared memory
systems is well studied in the context of lock acquisition and release.
When it comes to message passing, the problem of concurrent accesses to
\emph{channels}, in the absence of shared memory, is also well
studied---the goal, in these cases, is to achieve determinism rather than
race-freedom~\cite{cypher1995efficient, damodaran1993nondeterminancy,
  terauchi2008capability}.  What is less prominent in the race-detection
literature is the study of channel communication as the synchronization
primitive for shared memory systems.  In this paper, we present exactly
that; a \iftrace sound and complete \fi dynamic data-race detector for a
language in the style of Go, featuring channel communication as means of
coordinating accesses to shared memory.

We fix the syntax of our calculus in Section~\ref{sec:race.detector} and
present a corresponding operational semantics. The configurations of the
semantics keep track of memory events (i.e. of read and write accesses to
shared variables) such that the semantics can be used to detect races.  A
proper book-keeping of events also involves tracking
\emph{happens-before} information.  In the absence of a global clock, the
happens-before relation is a vehicle for reasoning about the relative
order of execution of different threads~\cite{lamport:time}.
We describe the race detection task and present a framework, called \grace~\cite{www:grace}, that is based on what we call \emph{happens-before sets}.  Different from other race 
detectors, which often employ vector clocks (VCs) as a mechanism for
capturing the happen-before relation, we tie our formalization more closely
to the concept of happens-before.
The proposed approach, based on what we call happens-before sets, allows
for garbage collection of ``stale'' memory access information that would
otherwise be tracked.
Although, in the worst case, the proposed detector requires a larger
foot-print when compared to VC-based implementations, we conjecture the
existence of a hybrid approach that can offer benefits from both worlds.

Our race detector is built upon a previous
result~\cite{fava*:oswmm-chan-journal}, where we formalize a weak memory
model inspired by the Go specification~\cite{www:gomemorymodel}.  The
core of the paper was a proof of the DRF-SC guarantee, meaning, we proved
that the proposed relaxed memory model behaves Sequentially Consistently
(SC) when running Data-Race Free (DRF) programs.  The proof hinges on the
fact that, in the absence of races, all threads agree on the contents of
memory.  The scaffolding used in the proof contains the ingredients for the
race detector presented in this paper.
We should point out, however, that the operational semantics presented here
and used for race detection is \emph{not} a weak semantics.%
\footnote{Note that while the mentioned semantics
  of~\cite{fava*:oswmm-chan-journal} differs from the one presented here,
  both share some commonalities. Both representations are based on
  appropriately recording information of previous read and write events in
  their run-time configuration. In both versions, a crucial ingredient of
  the book-keeping is connecting events in happens-before relation. The
  purpose of the book-keeping of events, however, is different:
  in~\cite{fava*:oswmm-chan-journal}, the happens-before relation serves to
  operationally formalize the weak memory model (corresponding roughly to
  PSO)\index{PSO} in the presence of channel communication.  In the current
  paper, the same relation serves to obtain a \iftrace sound and complete
  \fi race detector. Both versions of the semantics are connected by the
  DRF-SC result, as mentioned.}  Apart from the additional information for
race detection, the semantics is ``strong'' in that it formalizes a memory
guaranteeing \emph{sequential consistency.}  To focus on a form of strong
memory is not a limitation.  Since we have established that a corresponding
weak semantics enjoys the crucial DRF-SC\index{DRF-SC}
property~\cite{fava*:oswmm-chan-journal}, the strong and weak semantics
agree up to the first encountered race condition.  Given that even racy
program behaves sequentially consistently up to the point in which the
first data-race is encountered, a complete race detector can safely operate
under the assumption of sequential consistency.

\medskip

The remainder of the paper is organized as follows.  
Section~\ref{sec:race.background} presents background information on data races and synchronization via message passing that are directly related to the formalization of our approach to race detection.
Section~\ref{sec:race.detector} formalizes race detection in the context of channel communication as sole synchronization mechanism.
\iftrace
Section~\ref{sec:race.trace} introduces a trace grammar and defines data races on execution histories. An independence relation on events (technically on event labels) is defined which allows us to reason about equivalent histories.  Soundness and completeness of the data-race detector is then proven in Section~\ref{sec:race.correct} by relating the execution of a program in the operational semantics to histories according to the trace grammar.  
\fi
We turn our attention to the issue of efficiency in Section~\ref{sec:race.detector.gc}.  
Section~\ref{sec:race.comparison} gives a detailed comparison of our algorithm and VC-based algorithms for the acquire-release semantics.
Section~\ref{sec:race.discussion} puts our work in the perspective of trace theory.
Section~\ref{sec:race.related} examines related work.  Section~\ref{sec:race.conclusion} provides a conclusion and touches on future work.


%% file: background.tex
\section{Background}
\label{sec:race.background}

\paragraph{Read and write conflicts.}
\label{sec:race.background.conflicts}

Memory accesses conflict if they target the same location and at least one
of the accesses is a \emph{write}---there are no read-read conflicts.  A
data race constitutes of conflicting accesses that are unsynchronized.

\begin{figure}[thb]
  \centering
  \lstset{frame=single,language=go}
  \begin{lstlisting}[%numbers=left,numberstyle=\tiny,
      label={lst:race.ex.race},
      caption={Program with race condition. \cite{www:gomemorymodel}}]
var a string

func main() {
	go func() { a = "hello" }()
	print(a)
}
  \end{lstlisting}
\end{figure}

Take the Go code of Listing~\ref{lst:race.ex.race} as an example.
There, the main function invokes an anonymous function; this anonymous function sets the global variable ``\lstinline[basicstyle=\normalsize,language=go]{a}'' to ``\lstinline[basicstyle=\normalsize,language=go]{hello}''.
Note, however, that the call is prepended with the keyword \lstinline[basicstyle=\normalsize,language=go]{go}.  When this keyword is present in a function invocation, Go spawns a new thread (or goroutine), and the caller continues execution without waiting for the callee to return.
The main and the anonymous functions access the same shared variable in a conflicting manner (\textit{i.e.} one of the accesses is a write).
Since both the main and the anonymous functions run in parallel and no synchronization is used (as evidenced by the lack of channel communication), the two accesses are also concurrent.
This allows us to conclude that this program has a race.

A data race \emph{manifests} itself when an execution step is
immediately followed by another and the two steps are conflicting.  This
definition is the closest one can get to a notion of simultaneity in an
operational semantics, where memory interactions are modeled as
instantaneous atomic steps.  While manifest races are obvious and easy to
account for, races in general can involve accesses that are arbitrarily far
apart in a linear execution.  A ``memory-less'' detector can fail to report
races, for example non-manifest races, that could otherwise be flagged by
more sophisticated race detectors.  The ability to flag non-manifest
data-races is correlated with the amount of information kept and the length
in which this information is kept for.  In general, recording more
information and storing it for longer leads to higher degrees of
``completeness'' at the expense of higher run-time overheads.\footnote{It
  should go without saying that observing one execution as being race free
  is not enough to assert race-freedom of the program, even if one has
  observed a complete trace of a terminating run of a program. Completeness
  can at best be expected with respect to alternative schedules or
  linearizations of a given execution.}

We break down the notions of read-write and write-write conflicts into a more fine-grained distinction.  Inspired by the notion of data hazards in the computer architecture literature, we break down read-write conflicts into read-after-write (RaW) and write-after-read (WaR) conflicts.  To keep consistent with this nomenclature, we refer to write-write conflicts as write-after-write (WaW).%
\footnote{The mentioned ``temporal'' ordering and the use of the word
  ``after'' refers to the occurrence of events in the trace or execution of
  the running program. It is incorrect to conflate the concept of
  happens-before with the ordering of occurrences in a trace.  For
  instance, in a RaW situation, the read step occurs after a write in an
  execution, i.e., the read is mentioned after the write in the
  linearization. This order of occurrence does \emph{not} mean, however,
  that the read happens-after the write or, conversely, the write
  happens-before the read. Actually, for a RaW race (same as for the other
  kinds of races), the read occurs after the write but the accesses are
  \emph{concurrent}, which means that they are \emph{unordered} as far as
  the happens-before relation is concerned.}
Going back to the example in Listing~\ref{lst:race.ex.race}, there are two possible executions: one in which the spawned goroutine writes ``\lstinline[basicstyle=\normalsize,language=go]{hello}'' to the shared variable after the main function prints it, 
and another execution in which the print occurs after the writing of the variable.  The first execution illustrates a write-after-read race, while the first illustrates a read-after-write. Note that this example does not contain a write-after-write race.

We make the distinction between the detection of after-write races and the
detection of write-after-read ones.
As we will see in Section~\ref{sec:race.detector.raw.waw}, the detection of after-write races can be done with little overhead.  The detection of after-read, however, cannot.

When reading or writing a variable, it must be checked that conflicting
accesses \emph{happened-before} the current access.  The check must happen
from the perspective of the thread attempting the access.  In other words,
the question of whether an event occurred in the ``definite past'' (i.e.,
whether an event is in happened-before relation with ``now'') is
\emph{thread-local}; threads can have different views on whether an event
belongs to the past.  This thread-local nature is less surprising than it
may sound: if one threads executes two steps in sequence, the second step
can safely assume that the first has taken effect; after all, that is what
the programmer must have intended by sequentially composing instructions in
the given program order.  Such guarantees hold locally, which is to say
that the semantics \emph{respects program order within a thread}.  It is
possible, however, for steps to not take effect in program order.  A
compiler or hardware may rearrange instructions, and it often does so in
practice.  What must remain true is that these reorderings cannot be
observable from the perspective of a single thread.  When it comes to more
than one thread, however, agreement on what constitutes the past cannot be
achieved without synchronization.  Synchronization and consensus are
integrally related.\footnote{In the context of channel communication and
  weak memory, the connection between synchronization and consensus is
  discussed in a precise manner in our previous work; see the
  \emph{consensus lemmas} of~\cite{fava*:oswmm-chan-journal}.}
Specifically, given a thread $t$, events from a different thread $t'$ are
not in the past of $t$ unless synchronization forces them to be.

\paragraph{Synchronization via bounded channels.}
\label{sec:race.background.channels}
\index{bounded channel}

In the calculus presented here, channel communication is the only way in
which threads synchronize.  Channels can be created dynamically and closed; they are also first-class data, which means channel identifiers can be
passed as arguments, stored in variables, and sent over channels.  Send and
receive operations are central to synchronization.  Clearly, a receive
statement is synchronizing in that it is potentially blocking: a thread
blocks when attempting to receive from an empty channel until, if ever, a
value is made available by a sender.  Since channels here are bounded,
there is also potential for blocking when sending, namely, when attempting
to send on a channel that is full.\index{bounded channel}

We can use a channel \lstinline[basicstyle=\normalsize,language=go]{c} to eliminate the data race in Listing~\ref{lst:race.ex.race} as follows: the anonymous function sends a message to communicate that the shared variable has been set.  Meanwhile, the main thread receives from the channel before printing the shared variable.

\begin{figure}[thb]
  \centering
  \lstset{frame=single,language=go}
  \begin{lstlisting}[%numbers=left,numberstyle=\tiny,
      label={lst:race.ex.no.race},
      caption={Repaired program.}]
var a string
var c = make(chan bool, 1);

func main() {
	go func() { a = "hello";  c <- true }()
	<- c
	print(a)
}
  \end{lstlisting}
\end{figure}

The happens-before memory model stipulates, not surprisingly, a causal relationship between the communicating partners~\cite{www:gomemorymodel}:
\begin{eqnarray}
  \label{eq:gomm.hb.chan.forward}
  & & \text{A send on $c$ happens-before the corresponding receive from $c$ completes.}~~~~
\end{eqnarray}

Given that channels have finite capacity, a thread remains blocked when
sending on a full channel until, if ever, another process frees a slot in
the channel's buffer.  In other words, the sender is blocked until another
thread receives from the channel.  Correspondingly, there is a
happens-before relationship between a receive and a subsequent send on a
channel with capacity $k$~\cite{www:gomemorymodel}:
\begin{eqnarray}
  \label{eq:gomm.hb.chan.backward}
  & & \text{The $i^{\mathit{th}}$ receive from $c$ happens-before the $(i+k)^{\mathit{th}}$ send on $c$ completes.}
\end{eqnarray}
Interestingly, because of this rule, a causal connection is forged between
the sender and \emph{some previous} receiver who is otherwise unrelated to
the current send operation.  When multiple senders and receivers share a
channel, rule~(\ref{eq:gomm.hb.chan.backward}) implies that it is possible
for two threads to become related (via happens-before) without ever
directly exchanging a message.\footnote{Communication means sending a
  message to or receiving a message from a channel; messages are not
  addressed to or received from specific threads. Thus, sharing the channel
  by performing sends and receives does not necessarily make two threads
  ``communication partners.''  Two threads are partners when one receives a
  message deposited by the other.}

The indirect relation between a sender and a prior receiver, postulated by
rule~(\ref{eq:gomm.hb.chan.backward}), allows channels to be used as locks.
In fact, free and taken binary locks are analogous to empty and full
channels of capacity one.  A process takes and releases locks for the
purpose of synchronization (such as assuring mutually exclusive access to
shared data) without being aware of ``synchronization partners.''  In the
(mis-)use of channels as locks, there is also no inter-process
communication.  Instead, a process ``communicates'' with itself: In a
proper lock protocol, the process holding a lock (i.e. having performed a
send onto a channel) is the only one supposed to release the lock
(i.e. performing the corresponding receive).  Thus, a process using a
channel as lock receives its own previously sent message---there is no
direct inter-process exchange.  Note, however, synchronization still
occurs: subsequent accesses to a critical region are denied by sending onto
a channel and making it full.  See Section~\ref{sec:race.example.mutex} for
a more technical elaboration.

To establish a happens-before relation between sends and receives, note the
distinction, between a channel operation and its \emph{completion} in the
formulation of rules~(\ref{eq:gomm.hb.chan.forward}) and~(\ref{eq:gomm.hb.chan.backward}).
The order of events in a concurrent system is partial; not only that, it is strictly partial since we don't think of an event as happening-before itself.
A strict partial order is an irreflexive, transitive, and asymmetric relation.  In the case of synchronous channels, if we were to ignore the distinction between an event and its completion, according to rule~(\ref{eq:gomm.hb.chan.forward}), a send would then happen-before its corresponding receive, and, according to rule~(\ref{eq:gomm.hb.chan.backward}), the receive would happen-before the send.  This cycle breaks asymmetry.  Asymmetry can be repaired by interpreting a send/receive pair on a synchronous channel as a single operation; indeed, it can be interpreted as a \emph{rendezvous}.

The distinction between a channel operation and its completion is arguably
more impactful when it comes to buffered channels.  For one, it prevents
sends from being in happens-before with other sends, and receives from
being in happens-before with other receives.  To illustrate, let $\sds^i$
and $\rvs^i$ represent the $i^\mathit{th}$ send and receive on a channel.
If we remove from rules~(\ref{eq:gomm.hb.chan.forward})
and~(\ref{eq:gomm.hb.chan.backward}) the distinction between an operation and
its completion, the $i^\mathit{th}$ receive would then happens-before the
$(i+k)^\mathit{th}$ send---based on rule~(\ref{eq:gomm.hb.chan.backward})---and
the $(i+k)^\mathit{th}$ send would happens-before the $(i+k)^\mathit{th}$
receive---based on rule~(\ref{eq:gomm.hb.chan.forward}):
\begin{displaymath}
  \rvs^i ~\hbarrow~ \sds^{i+k} ~\hbarrow~ \rvs^{i+k}
\end{displaymath}
By transitivity of the happens-before relation, we would then conclude that
the $i^\mathit{th}$ receive happens-before the $(i+k)^\mathit{th}$ receive,
which would happen-before the $(i+2k)^\mathit{th}$ receive and so on.  As a
consequence, a receive operation would have a lingering effect through-out
the execution of the program---similarly for send operations.  This
accumulation of effects can be counter intuitive for the application
programmer, who would be forced to reason about arbitrarily long histories.

%% file: race.tex
\section{Data-race detection}
\label{sec:race.detector}

We start in Section~\ref{sec:race.calculus} by presenting the abstract syntax of our calculus and, in Section~\ref{sec:race.detector.overview}, an overview of the operational semantics used for data-race detection.
The race detector itself is introduced incrementally.
We start in Section~\ref{sec:race.detector.raw.waw} with a simple detector that has a small footprint but that is limited to detecting after-write races.
We build onto this first iteration of the detector in Section~\ref{sec:race.detector.war}, making it capable of detecting after-write as well as after-read races.
The detector's operation is illustrated by examples in Section~\ref{sec:race.example}.
Later, in Section~\ref{sec:race.detector.gc}, we turn to the issue of efficiency and introduce ``garbage collection'' as a mean to reduce the detector's footprint.
These race detectors can be seen as augmented versions of an underlying semantics without additional book-keeping related to race checking.  This ``undecorated'' semantics, including the definition of internal steps and a notion of structural congruence, can be found in Appendix~\ref{sec:gomm.race.semantics.strong}.

\subsection{A  calculus with shared variables and channel communication}
\label{sec:race.calculus}

\index{Go}We formalize our ideas in terms of an idealized language shown in
Figure~\ref{fig:race.grammar} and inspired by the Go programming
language.
\begin{figure}[thb]
  \centering
  \input{rules/grammar}
  \caption{Abstract syntax\label{fig:race.grammar}}
\end{figure}
The syntax is basically unchanged from
\cite{fava*:oswmm-chan-journal}. \index{value}\index{channel name}%
\emph{Values} $v$ can be of two forms: $r$ denotes local variables or
registers; $n$ is used to denote references or names in general and,
in specific, $p$ for processes or goroutines, $m$ for memory events, and
$c$ for channel names. We do not explicitly list values such as the unit
value, booleans, integers, etc.\index{local variable}\index{$r$ (local
  variable)} We also omit compound local expressions like $e_1 + e_2$.
Shared variables are denoted by $x$, $z$, etc., $\load{z}$\index{load
  z@$\load{z}$} represents reading the shared variable $z$ into the thread,
and $\store{z}{v}$ denotes writing to $z$.%
\index{$\store{z}{v}$} References are dynamically created%
\iflong
~and are, therefore, part of the \emph{run-time} syntax.  Run-time syntax is highlighted in the grammar with an underline as in $\rts{n}$. \index{run-time syntax}\index{syntax!run-time}
\else
.
\fi
A new channel is
created by $\makechans{T}{v}$\index{make@$\makechans{T}{v}$}, where $T$
represents the type of values carried by the channel and $v$ a non-negative
integer specifying the channel's capacity. Sending a value $v$ over a channel
$c$ and receiving a value as input from a channel are denoted respectively as
$\gsend{c}{v}$ and $\greceive{c}$. %
\index{$\gsend{c}{v}$ (send)}\index{$\greceive{c}$ (receive)} After the
operation $\gcloses$, no further values can be sent on the specified
channel.  Attempting to send values on a closed channel leads to a
panic.\index{panic} \index{close a channel}

Starting a new asynchronous activity, called goroutine in Go, is done using
the $\gos$-keyword. In \Go, the $\gos$-statement is applied to function
calls only. We omit function calls, asynchronous or otherwise, as they are
orthogonal to the memory model's formalization. The select-statement, here
written using the $\selects$-symbol, consists of a finite set of branches
(or communication clauses in Go-terminology).  These branches act as
guarded threads.  General expressions in Go can serve as guards.  Our
syntax requires that only communication statements (i.e., channel sending
and receiving) and the $\defaults$-keyword can serve as guards.  This does
not reduce expressivity and corresponds to an\index{A-normal form} A-normal
form representation~\cite{sabry.felleisen.continuation-proc}.
At most one branch is guarded by $\defaults$ in each select-statement.  The
same channel can be mentioned in more than one guard.  ``Mixed
choices''~\cite{palamidessi:comparing,
  peters.nestmann:isitgood}\index{choice!mixed}\index{mixed choice} are
also allowed, meaning that sending- and receiving-guards can both be used
in the same select-statement.  We use $\stops$ as syntactic sugar for the
empty select statement;\index{stop@$\stops$} it represents a permanently
blocked thread.  The $\stops$-thread is also the only way to syntactically
``terminate'' a thread, meaning that it is the only element of $t$ without
syntactic sub-terms.

The $\lets$-construct $\lets r = e \ins t$ combines sequential composition
and scoping for local variables $r$.  After evaluating $e$, the rest $t$ is
evaluated where the resulting value of $e$ is handed over using $r$. The
let-construct acts as a binder for variable $r$ in $t$. When $r$ does not
occur free in $t$, $\lets$ boils down to \index{let@$\lets$}%
\index{sequential composition}\emph{sequential composition} and, therefore,
is more conveniently written with a semicolon. See also
Figure~\ref{fig:race.syn.sugar} in the appendix for syntactic sugar.

\subsection{Overview of the operational semantics}
\label{sec:race.detector.overview}

To capture the notion of ordering of events between threads, an otherwise unadorned operational semantics (equation~(\ref{eq:gomm.race.naked.configs})) is equipped with additional information:  each thread and memory location tracks the events it is aware of as having happened-before---see the happens-before set $\idshb$ in the run-time configurations of equation~(\ref{eq:gomm.race.configs-simple}) and~(\ref{eq:gomm.race.gc.configs}), this set is present in terms corresponding to threads, $\ngoroutine{p}{\idshb}{t}$, as well as memory locations, $\ezwritesl{z}{v}{\idshbz{}}$ or $\erwritesl{m}{z}{v}{\idshbr}$.
Depending on the capabilities of the race detector, slightly different information is tracked as having happened-before (i.e. stored in a happens-before set).

\subsubsection{After-write races}

When detecting after-write races (i.e. RaW and WaW), in order to know
whether a subsequent access to the same variable occurs without proper
synchronization, one has to remember additional information concerning past
write-events.  Specifically, it must be checked that all write events to
the same variable \emph{happened-before} the current access.  The
happens-before set is then used to store information pertaining to write
events; read events are not tracked.  Also, terms representing a memory
location have a different shape when compared to the undecorated semantics.
In the undecorated semantics, the content $v$ of a variable $z$ is written
as a pair $\ewritesl{}{z}{v}$.  When after-write races come into play, it
is not enough to store the last value written to each variable; we also
need to identify write events associated with the variable.  Thus, an entry
in memory takes the form $\ezwritesl{z}{v}{\idshb}$ where $\idshb$ holds
identifiers $m$, $m'$, etc. that uniquely identify write events to
$z$---contrast the run-time configurations in
equation~(\ref{eq:gomm.race.naked.configs})
and~(\ref{eq:gomm.race.configs-simple}).  The number of prior write events
that need to be tracked can be reduced for the sake of efficiency, in which
case the term representing a memory location takes the form
$\erwritesl{m}{z}{v}{\idshbr}$ where $m$ is the identifier of the most
recent write to $z$.  See equation~(\ref{eq:gomm.race.gc.configs}).

\subsubsection{Write-after-read races}
Besides the detailed coverage of RaW and WaW races in
Section~\ref{sec:race.detector.raw.waw}, we describe the detection of
\emph{write-after-read} races in Section~\ref{sec:race.detector.war}.  When
it comes to WaR, the race checker needs to remember information about past
reads in addition to past write events.  Abstractly, a read event
represents the fact that a load-statement has executed.  Thus, the set
$\idshb$ of an entry $\ezwritesl{z}{v}{\idshb}$ in memory holds identifiers
of both read and write events.

In the strong semantics, a read always observes one definite value which is
the result of one particular write event.  Therefore, the configuration
contains entries of the form $\erwritesl{m}{z}{v}{\idshbr}$ where $m$ is
the identifier of the ``last'' write event and $\idshbr$ is a set of
identifiers of read events, namely those that accumulated after $m$.  Note
that ``records'' of the form $\erwritesl{m}{z}{v}{\idshbr}$ can be seen as
$n+1$ recorded events, one write event together with $n \geq 0$
read-events.  \ifreport Writing the recorded read events as in
equation~(\ref{eq:race.parallelreadevents}), the notation
$\erwritesl{m}{z}{v}{\idshbr}$ is a compact short-hand for
$\ewritesl{m}{z}{v_1} \parallel \ereadfrom{r_1}{m} \parallel
\ereadfrom{r_2}{m} \parallel \ldots$. \fi This definition of records with
one write per variable stands in contrast to a weak semantics, where many
different write events may be observable by a given
read~\cite{fava*:oswmm-chan-journal}.

\medskip

\subsubsection{Synchronization}
\label{sec:race.detector.channels}
\index{bounded channel}

Channel communication propagates happens-before information between threads, and thus, affects synchronization.
In the operational rules, each channel $c$ is actually realized with
\emph{two} channels, which we refer to as \emph{forward}, $c_f$, and
\emph{backward}, $c_b$---see Figure~\ref{fig:race.steps.chans.full}.
\index{forward channel}%
\index{backward channel}%
\index{channel!forward}%
\index{channel!backward}%
The forward part serves to communicate a value transmitted from a sender to a receiver; it also stipulates a causal relationship between the communicating partners~\cite{www:gomemorymodel}---see rule~(\ref{eq:gomm.hb.chan.forward}) of page~\pageref{eq:gomm.hb.chan.forward}.
To capture this relationship in the context of race checking, the sender
also communicates its current information about the happens-before relation
to the receiver.  The communication of happens-before information is
accomplished by the transmission of $\idshb$ over channels; see
rule~\rn{R-Rec} in Figure~\ref{fig:race.steps.chans.full}.

The memory model also stipulates a happens-before relationship between a receive and a subsequent send on a channel with capacity $k$---see rule~(\ref{eq:gomm.hb.chan.backward}) of page~\pageref{eq:gomm.hb.chan.backward}.
While we refer to the \emph{forward channel} as carrying a message from a sender to a receiver, the backward part of the channel is used to model the indirect connection between some prior receiver and a current sender; see \rn{R-Send} in Figure~\ref{fig:race.steps.chans.full}.

The interplay between forward and backward channels can also be understood
as a form of \index{flow control}flow control.  Entries in the backward
channel's queue are not values deposited by threads.  Instead, they can be
seen as tickets that grant senders a free slot in the communication
channel, i.e., the forward channel.%
\footnote{In the case of lossy channels, backward channels are sometimes
  used for the purpose of error control and regulating message
  retransmissions, where the receiver of messages informs the sender about
  the successful or also non-successful reception of a message.  Here,
  channels are assumed non-lossy and there is no need for error
  control.\index{error control} In that sense, the term ``backward'' should
  not be interpreted as communication \emph{back} to the receiver in the
  form of an acknowledgment.}  Thus, the number of ``messages'' in the
backward channel capture the notion of fullness: a channel is full if the
backward channel is empty.  See rule~\rn{R-Send} in
Figure~\ref{fig:race.steps.chans.full} or
Figure~\ref{fig:race.steps.chans.naked} for the underlying semantics
without race checking.
When a channel of capacity $k$ is created, the forward queue is empty and the backward queue is initialized so that it contains dummy elements $\lstateempty$ (cf.\ rule~\rn{R-Make}).  The dummy elements represent the number of empty or free slots in the channel. Upon creation, the number of dummy elements equals the capacity of the channel.

As discussed in Section~\ref{sec:race.background}, there is a distinction between a synchronization operation and its completion.
A send/receive pair on a synchronous channel can be seen as a rendezvous operation; captured in our semantics by the \rn{R-Rend}~reduction rule of Figure~\ref{fig:race.steps.chans.full}.
When it comes to asynchronous communication, the distinction between a channel operation and its completion is handled by the fact that send and receive operations update a thread's local state but do not immediately transmit the updated state onto the channel---see rules~\rn{R-Send} and~\rn{R-Rec} in Figure~\ref{fig:race.steps.chans.full}.

{
  \def\VERSION{.semi}
  \input{race-semi}
}

{
  \def\VERSION{.full}
  \input{race-full}

}

\subsection{Examples}
\label{sec:race.example}

We will look at two examples of properly synchronized programs.  The first
is a typical usage of channel communication; one in which an action is placed
in the past of another.  The second example relies on mutual exclusion
instead.  In this case, we know that actions are not concurrent, but we cannot
infer an order between them.
By contrasting the two examples in Section~\ref{sec:race.determinism}, we derive observations related to determinism and constructivism.

\subsubsection{Message passing}
\label{sec:race.example.mp}
Message passing, depicted in Figure~\ref{fig:ex.message.passing}, involves a producer writing to a shared variable and notifying another thread by sending a message onto a channel.  A consumer receives from the channel and reads from the shared variable.

\begin{figure}[thb]
\begin{align*}
  \qquad\qquad\qquad\qquad\qquad\qquad
  &\ngoroutine{p_1}{{\idshb}_1}{\store{z}{42};\ \gsend{c}{0}} \\
  &\ngoroutine{p_2}{{\idshb}_2}{\greceive{c};\ \load{z}}
\end{align*}
\caption{Message passing example. \label{fig:ex.message.passing}}
\end{figure}

The access to the shared variable is properly synchronized.  Given the operational semantics presented in this chapter, we can arrive at this conclusion as follows.
A fresh label, say $m$, is generated when $p_1$ writes to $z$.  The memory record involving $z$ is updated with this fresh label, and the pair $\ewritea{m}{z}$ is placed into $p_1$'s happens-before set, thus yielding ${\idshb}_1'$.
A send onto $c$ sends not only the message value, $0$ in this case, but also the happens-before set of the sender, ${\idshb}_1'$, see rule~\rn{R-Send}.
The act of receiving from $c$ blocks until a message is available.  When a message becomes available, the receiving thread receives not only a value but also the happens-before set of the sender at the time that the send took place, see rule~\rn{R-Rec}.
Thus, upon receiving from $c$, $p_2$'s happens-before set is updated to contain $\ewritea{m}{z}$. Receiving from the channel places the writing to $z$ by $p_1$ into $p_2$'s definite past. 
The race-checker makes sure of this fact by inspecting $p_2$'s happens-before set when $p_2$ attempts to load from $z$.
In other words, the race-checker checks that the current labels associated with $z$ in the configuration are also present in the happens-before set of the thread performing the load.

The message passing example illustrates synchronization as imposing of an order between events belonging to different threads.  The message places the producer's write in the past of the consumer's read.
Next, we will look into an example in which synchronization is achieve via mutual exclusion.  Two threads, $p_1$ and $p_2$, are competing to write to the same variable.  We will not be able to determine which write happens-before the other.  Even though we cannot infer the order, we can determine that a happens-before order exists and, therefore, that the program is properly synchronized.

\subsubsection{Mutual exclusion}
\label{sec:race.example.mutex}

Figure~\ref{fig:ex.mutual.exclusion} shows a typical mutual exclusion
scenario.  It involves two threads writing to a shared variable $z$.
Before writing, a thread sends a message onto a channel $c$ which capacity
$\len{c}=1$.  After writing, it receives from $c$.%
\footnote{Note that the channel is being used as a semaphore~\cite{dijkstra:over}.  Sending on the channel is analogous to a semaphore \texttt{wait} or \texttt{P} operation.  Receive is analogous to \texttt{signal} or \texttt{V}.  
The \texttt{wait} decrements the value of the semaphore and, if the new value is negative, the process executing the \texttt{wait} is blocked.  A \texttt{signal} increments the value of the semaphore variable, thus allowing another process (potentially coming from the pool of previously blocked processes) to resume.
Similarly, a send operation decrements the number of available slots in the
channel's queue, while a receive increments it.  Sending on a channel with
capacity 1 can only take place if the channel is empty; meaning, all
previous sends are matched with a corresponding receive.}

\begin{figure}[thb]
\begin{align*}
  \qquad\qquad\qquad\qquad\qquad\qquad
  &\nslgoroutine{p_1}{\gsend{c}{0};\ \store{z}{17};\ \greceive{c}} \\
  &\nslgoroutine{p_2}{\gsend{c}{0};\ \store{z}{42};\ \greceive{c}}
\end{align*}
\caption{Mutual exclusion example. \label{fig:ex.mutual.exclusion}}
\end{figure}

A send and its corresponding receive do not directly contribute to synchronization in this example.  The send is matched by a receive from the same thread; nothing new is learned from this exchange.
To illustrate this point, which may come as a surprise, let us look at an execution.  Say $p_1$ is the first to send $0$ onto $c$.  Then $p_1$'s happens-before set ${\idshb}_1$ is placed onto the channel along with the value of $0$.  The thread then proceeds to write to $z$, which generates a fresh label, say $m'$; the pair $\ewritea{m'}{z}$ is placed on $p_1$'s happens-before set.
When receiving from $c$, $p_1$ does not learn anything new!
It receives the message $0$ and a ``stale'' happens-before set ${\idshb}_1$. 
The receiver's happens-before set, ${\idshb}_1'$, is updated to incorporate the  stale happens-before set, but this ``update'' causes no effective change:
\begin{align*}
{\idshb}_1' \cup {\idshb}_1 &= 
  ({\idshb}_1 \cup \{ \ewritea{m'}{z} \}) \cup {\idshb}_1 \\
  &= {\idshb}_1 \cup \{ \ewritea{m'}{z} \} \\
  &= {\idshb}_1'
\end{align*}

The explanation for why the program is synchronized, in this case, is more subtle.  It involves reasoning about the channel's capacity.
Recall that, according to rule~(\ref{eq:gomm.hb.chan.backward}) on page~\pageref{eq:gomm.hb.chan.backward}, the $i^\mathit{th}$ receive from a channel with capacity $k$ happens before the $(i+k)^\mathit{th}$ send onto the channel completes.
Since channel capacity is $1$ in our example, rule~(\ref{eq:gomm.hb.chan.backward}) implies that the first receive from the channel happens-before the second send completes.  If $p_1$ is the first to write to $z$, then $p_1$ is also the first to receive from $c$.  Receiving from $c$ places $p_1$'s happens-before set onto the backward channel (see rule~\rn{R-Rec}).  This happens-before set contains the entry $\ewritea{m'}{z}$ registering $p_1$'s write to $z$.
Upon \emph{sending} onto $c$, $p_2$ receives from the backward channel and learns of $p_1$'s previous write.
Thus, by the time $p_2$ writes to $z$, the write by $p_1$ has been-placed onto $p_2$'s definite past.  Since no concurrent accesses exist, the race checker does not flag this execution as racy.

Similarly, $p_2$ could first send onto $c$ and write to $z$.  The argument for the proper synchronization of this alternate run would proceed in the same way.
Therefore, even though it is not possible to infer who, among $p_1$ and $p_2$, writes to $z$ first, we know that one of the writes is in a happens-before relation with the other.  This knowledge is enough for us to conclude that the program is properly synchronized.

This example shows that channels are excessively powerful when it comes to implementing mutual exclusion, as evidenced by the fact that the forward queue associated with the channel is not utilized.  When it comes to mutual exclusion, a more parsimonious synchronization mechanism suffices.  Indeed, the \emph{acquire} and \emph{release} semantics associated with locks is a perfect fit.
When acquiring a lock, a thread \textit{learns} about the memory operations that precede the lock's release.  In other words, memory operations preceding a lock's release are put in happens-before with respect to a thread that acquires the lock.  
Assuming a lock $l$ starts with empty happens-before information, say $\tlock{l}{\emptyset}$, the rules \rn{Acquire} and \rn{Release} capture a lock's behavior.

\begin{ruleset}
\infrule{Acquire}{
  \idshb' = \idshb \cup \idshb''
}{
  \ngoroutine{p}{\idshb}{\acq{l};t} \parallel
  \tlock{l}{\idshb''}
  \widetrans{}
  \ngoroutine{p}{\idshb'}{t} \parallel
  \tlock{l}{}
}

\ruleskip

\infrule{Release}{
  \idshb' = \idshb \cup \idshb''
}{
  \ngoroutine{p}{\idshb}{\rel{l};t} \parallel
  \tlock{l}{}
  \widetrans{}
  \ngoroutine{p}{\idshb}{t} \parallel
  \tlock{l}{\idshb'}
}
\end{ruleset}

\medskip

Note that an acquired lock, represented by $\tlock{l}{}$, cannot be re-acquired without a prior release, and that a released lock, meaning $\tlock{l}{\idshb}$, cannot re-released without a prior acquire.%
\footnote{When releases are matched by an prior acquire from the same thread, then happens-before information accumulates monotonically, meaning, a thread learns about all previous releases, not just the most recently occurring one.}
While a thread's happens-before is updated on both sends and receives, with locks, only the acquisition updates a thread's happens-before information.
Surrounding code with a call to acquire at the beginning and release at the end is sufficient for ensuring mutual exclusion.  The full generality of channels is not required.

\subsubsection{Determinism, confluence, and synchronization}
\label{sec:race.determinism}
In the message passing example of Section~\ref{sec:race.example.mp}, we are able to give a constructive proof-sketch of the synchronization between $p_1$ and $p_2$; the ``proof'' puts an event from $p_1$ in the past of $p_2$.
In the mutual exclusion example of Section~\ref{sec:race.example.mutex}, no such guarantee is possible.  Instead, we give a non-constructive ``proof'' that $p_1$ and $p_2$ are synchronized by arguing that either $p_1$'s actions are in the past of $p_2$'s or vice versa.  The \emph{law of excluded middle} is used in this non-constructive argument.

The absence of constructivism is tied to the absence of determinism.  While in the message passing example the program is deterministic, in the mutual exclusion example it is not.  There is no \emph{data} race in the mutual exclusion example, but there is still a ``race'' insofar as the two threads compete for access to a shared resource.  The resource, in this case, is the channel, which is being used as a lock.  The two threads race towards acquiring the lock (i.e. sending onto the channel) first.
The initial configuration has two transitions, one in which $p_1$ acquires the lock first and one in which $p_2$ does.  These transitions are non-confluent.

When it comes to reasoning about programs that model hardware, the lack of constructivism and the non-confluence in the use of channels as locks is a hindrance.  Deterministic languages and constructive logics are needed in order to rule out scenarios in which two logic gates attempt to drive the same \emph{via} with different logic values (i.e. a short circuit)~\cite{benveniste2003synchronous}.
In the case of channel communication and in the absence of shared memory, determinism can be achieved by enforcing ownership on channels; for example, by making sure a single thread can read and a single thread can write on a given channel at any given point in the execution~\cite{steffen.nestmann:typing-rep-doesthisexisted}.  It is possible for the ownership on channels to be passed around the threads in a way that preserves determinism~\cite{terauchi2008capability}.

The examples show that the absence of absence of data races is not enough to ensure determinism.  In general, however, determinism is not a requirement.  Many applications require ``only'' data-race freedom.


%% file: rules/grammar.tex
\begin{displaymath}
  \begin{array}[t]{rcl@{\quad}l}
    v  & \bnfdef & r \bnfbar \rts{n}  & \text{values}
    \\ 
    e& \bnfdef &  t \bnfbar 
    v
    \bnfbar
    \load{z}
    \bnfbar
    \store{z}{v}
    \bnfbar
    \gos t
    &
    \text{expressions}
    \\
    & \bnfbar & \ifs v \thens  t \elses t
    \\
    &  \bnfbar  &
    \makechans{\T}{v}
    \bnfbar
    \greceive{v}
    \bnfbar 
    \gsend{v}{v}
    \bnfbar 
    \gcloses v &
    \\
    g & \bnfdef & \gsendguard{v}{v}
    \bnfbar
    \greceiveguard{v}\bnfbar
    \bnfbar
    \defaults
& \text{guards}
    \\
    t & \bnfdef &
    \lets r = e \ins t
    \bnfbar
    \selects_i \lets r_i = g_i \ins  t_i   & \text{threads}
  \end{array}
\end{displaymath}
%

%% file: race-semi.tex
\subsection{Detecting read-after-write (RaW) and write-after-write (WaW) races}
\label{sec:race.detector.raw.waw}

\index{run-time configuration} \index{$R$ (run-time configuration)} To detect ``after-write'' races, run-time configurations are given following syntax:
\index{R@$R$ (run-time configuration)}

\input{./rules/datarace/configs-simple} 
\index{$\ezwritesl{z}{v}{\idshb}$}
Configurations are considered up-to structural congruence, with the empty
configuration $\emptyconf$ as neutral element and $\parallel$ as
associative and commutative.  The definition is standard and included in
Appendix~\ref{sec:gomm.race.cong}. Likewise relegated to the appendix are
\emph{local} reduction rules, i.e., those not referring to shared variables
or channels (see Appendix~\ref{sec:gomm.race.steps.local}). %
\index{$\emptyconf$ (empty configuration)}%
\index{structural congruence}%

\index{memory} \index{$\ewritesl{m}{z}{v}$ ((recorded) write event)} In the
configurations, a triple $\ezwritesl{z}{v}{\idshbz{z}}$ not only stores the
current value of $z$ but also records the unique identifiers $m$, $m'$, etc
of every write \emph{event} to $z$ in $\idshbz{z}$.%
\footnote{We will later use the term ``event'' also when talking about
  histories or traces.  There, events carry slightly different
  information. For instance, being interested in the question whether a
  history contains evidence of a race, it won't be necessary to mention the
  actual value being written in the write event in the history. Both
  notions of events, of course, hang closely together.  It should be clear
  from the context whether we are referring to events as part of a linear
  history or recorded as part of the configuration.  When being precise, we
  refer to a configuration event as \emph{recorded} event. Since recorded
  events in the semantics are uniquely labeled, we also allow ourselves to
  use words like ``event $m$'' even if $m$ is just the identifier for the
  recorded event $\ewritesl{m}{z}{v}$.}  A write to memory updates a
variable's value and also generates a fresh identifier $m$.  In order to
record the write event, the tuple $\ewritea{m}{z}$ is placed in the
happens-before set of the term representing the memory location that has
been written to.  The initial configuration starts with one write-event per
variable and the semantics maintains this uniqueness as an invariant.  In
effect, the collection of recorded write events behave as a mapping from
variable to values.%
\footnote{The fact that memory behaves like a mapping is consistent with
  the strong memory assumption.}

A thread $t$ is represented as $\ngoroutine{p}{\idshb}{t}$ at run-time, with $p$ serving as identifier.  To be able to determine whether a next action should be flagged as race or not, a goroutine keeps track of happens-before information corresponding to past write events. An event mentioned in $\idshb$ is an event of the past, as opposed to being an event that simply occurred in a prior step.
An event is ``concurrent'' if it occurred in a prior step but is not in happens-before relation with the current thread state.  Concurrent memory events are potentially in conflict with a thread's next step.
More precisely, if the memory record $\ezwritesl{z}{v}{\idshbz{z}}$ is part of the configuration, then it is safe for thread $\ngoroutine{p}{\idshb}{t}$ to write to $z$ if $\idshbz{z} \subseteq \idshb$.  Otherwise, there exist a write to $z$ that is not accounted for by thread $p$ and a WaW conflict is raised.  Similar when reading from a variable.

Data-races are marked as a transition to an exception $\Texc$---see the derivation rules of Figure~\ref{fig:race.steps.smem.excep.semi}, and, when write-after-read races are considered, Figure~\ref{fig:race.steps.smem.excep.full}. 
The exception takes as argument a set containing the prior memory operations that conflict and are concurrent with the attempted memory access.

\ifprivate

\begin{remark}[Updating the memory, weak semantics and race checker]
  \index{memory} As explained, the memory here associates values with
  variables. Actually, at each point in time, it associates to each shared
  variable exactly one value. In addition to that, it remembers a unique
  identifier, which makes (recorded) write events into triples
  $\ewritesl{w_1}{z}{v_1}$, where $w_1$ is the name of the event, $z$ the
  shared variable and $v_1$ the value. When writing to that variable, say a
  value $v_2$ in an event, say $w_2$, the above triple is replaced by
  $\ewritesl{w_2}{z}{v_2}$. That represents the conventional view of a
  global shared memory, which stores the latest value for each variable,
  here additionally decorated with identification of the responsible write,
  and were writing means replacing that variable (hand in hand with
  updating the event identifier). As straightforward as that is, it is a
  picture tightly connected with the notion of a strong memory model, a
  memory as global mapping from variables to values, which can be seen as a
  special case of more general, relaxed forms of memory. We can see triples
  $\ewritesl{w_1}{z}{v_1}$ also as (remembrance of) write
  \emph{events}. Events are generally understood as unique occurrences of
  ``steps'' or ``interactions'' occurring at run time, where here we are
  currently focusing on write interactions with the memory (later also of
  read interactions or channel interactions). In this point of view, a
  newer write event to a variable does not rescind the fact that an earlier
  write event had happened, it may only be no longer relevant. So instead
  of \emph{updating} value $v_1$ by $v_2$ (and likewise $w_1$ to $w_2$),
  one may also take the perspective, that the configuration keeps the
  record for \emph{two} write events:
  \begin{equation}
    \label{eq:config.parallelwriteevents}
    \ewritesl{w_1}{z}{v_1} \parallel     \ewritesl{w_2}{z}{v_2} \ ,
  \end{equation}
  where the second one occurred in a step after the first.  In a strong
  memory model, only the second write event is, at that point, relevant as
  it carries the legitimate value $v_2$, which is the legitimate value for
  \emph{all} threads (and $w_2$ the corresponding identifier). As
  additionally $w_1$, once out of date, will \emph{never} become relevant
  again, there is no need to keep it ``in memory'', so one may well
  garbage-collect it.  In weaker memory models, more than one value may
  currently be observable for a given variable, more than one write event
  needs to be kept in a configuration, and that executing a new write does
  not allow immediately to forget the previous one. As a consequence, a
  configuration may contain 2 or more write events, as sketched in equation
  (\ref{eq:config.parallelwriteevents}).

  \emph{Remembering} past (write) events is only half of the function a
  memory provides. Besides that, there have to be mechanisms to
  \emph{``forget''} outdated writes. In the strong memory model, it is the
  familiar overwriting of a value and replacing it by a new, for all
  threads instantaneously and globally. In a more relaxed memory model,
  different threads may a different views on which write events (and thus
  which values) are currently observable and which no longer. Such a
  ``disagreement'' does not exist in the strong semantics: there is always
  exactly one write event that is still visible, and it is the same for all
  threads (hence the memory needs to keep only the one and only consensus
  value).

  Such a \emph{consensus} on one-and-only observable write event is also a
  crucial invariant in the weak semantics for \emph{race-free} programs and
  at the heart of the so-called DRF-SC guarantee for weak memories: for
  race-free programs, the behavior under weak semantics coincides with the
  behavior under the strong semantics, i.e., it behaves sequentially
  consistent.

  The above statement concerning invariant for race-free programs under the
  weak semantics, namely that there is a consensus about the current value
  for each variable (represented by the consensus about the only observable
  corresponding write event) might need some refinement or
  elaboration. It is \emph{not} to be misunderstood that, at each time and
  for all processes, there is exactly one observable common write (as is
  the case in the strong semantics). The consensus on a common write is
  required \emph{only} for those processes and positions where an
  observation is actually done (i.e., a read executed). In other words:
  there is no need for a process that does \emph{not} execute a read-step
  on a variable, to be part, at that point, of a consensus about the
  currently valid value. There may as well be more than one write event
  observable for that process at that point, it only does not matter, as it
  does not do the observation by doing a read. The break of the consensus
  by doing that observation amounts to a race condition (in particular a
  read-after-write conflict). This shows the close link between the race
  checker, using a strong semantics, and the weak memory semantics, both
  based on a happens-before relation between events.
  \qed
\end{remark}

\fi

\ifprivate
\begin{remark}[Form of $\ewritea{m}{z}$ and $\idshb$]
  In connection with the weak semantics of~\cite{fava*:oswmm-chan-journal},
  there had been discussion about the form of $\idshb$, In particular
  whether it should mention the shared variable or not. In~\cite{fava*:oswmm-chan-journal}, it was decided to \emph{include} the
  variable, but it was also clear that technically it was not needed. Since
  the identifiers are unique, the variable is determined by the identifier
  alone. The reason why the variable was part of the information in
  $\idshb$ was for convenience or economy of formulation in the rules. In
  particular, it concerned the operational rule for writing. Unlike the
  semantics here, the weak semantics keep track of the identities of the
  ``shadowed writes'', and the rule \rn{R-Write} needed update those. In
  particular, it is convenient to maintain the ``shadow set''. When doing a
  write, all earlier write events to \emph{that} variable become outdated
  (``shadowed'').  As discussed elsewhere, very generally we can see it
  that the pair $\lstate$ is an abstraction of the
  ``happen-before-graph''. It is the relation between events (of which
  currently only the write-events are relevant) and to be self-contained,
  the events need to carry also the concerned variable.
  \qed%
\end{remark}
\fi

\begin{figure}[!ht]
  \centering
  \begin{ruleset}
    \input{rules/datarace/steps-smem-raw-waw-simple}
  \end{ruleset}
  \caption{Operational semantics augmented for RaW and WaW race detection\label{fig:race.steps.smem.semi.simple}}
\end{figure}

\begin{figure}[!ht]
  \centering
  \begin{ruleset}
    \input{rules/datarace/steps-smem-excep-simple}
  \end{ruleset}
  \caption{Exception conditions for RaW and WaW data-race detection\label{fig:race.steps.smem.excep.semi}}
\end{figure}

Goroutines synchronize via message passing, which means that channel communication must transfer happens-before information between goroutines. 
Suppose a goroutine $p$ has just updated variable $z$ thus generating the unique label $m$.  The tuple $\ewritea{m}{z}$ is placed in the happens-before set of both the thread $p$ and the memory record associated with $z$.
At this point, $p$ is the only goroutine whose happens-before set contains the label $m$ associated with this write-record.  No other goroutine can read or write to $z$ without causing a data-race.  When $p$ sends a message onto a channel, the information about $m$ is also sent.
Suppose now that a thread $p'$ reads from the channel and receives the corresponding message before $p$ makes any further modifications to $z$.
The tuple $\ewritea{m}{z}$ is added to $p'$'s happens-before set, so both $p$ and $p'$ are aware of $z$'s most recent write to $z$.
The existence of $m$ in both goroutine's happens-before sets implies that either $p$ or $p'$ are allowed to update $z$'s value.
The rules for channel communication are given in Figure~\ref{fig:race.steps.chans.full}. They will remain unchanged when we extend the treatment to RaW conflicts.  The exchange of happens-before information via channel communication is also analogous to the treatment of the weak semantics in~\cite{fava*:oswmm-chan-journal}.

\begin{figure}[!ht]
  \centering
  \begin{ruleset}
    \input{rules/datarace/steps-chans}
  \end{ruleset}
  \caption{Operational semantics augmented for race detection: channel communication\label{fig:race.steps.chans.full}} 
\end{figure}

As in \citet{fava*:oswmm-chan-journal}, ``the \rn{R-Close}~rule closes both sync and async channels.  Executing a receive on a \emph{closed} channel results in receiving the end-of-trans\-mission marker $\eot$ (cf.\ rule~\rn{R-Rec$_\eot$}) and updating the local state $\lstate$ in the same way as when receiving a properly sent value. \iflong This happens regardless of whether the channel is synchronous or not.\fi\ The ``value'' $\eot$ is not removed from the queue, so that all clients attempting to receive from the closed channel obtain the communicated happens-before synchronization information.''

Finally, goroutine creation is a synchronizing operation where the child, who is given a unique identifier $p'$, inherits the happens-before set from the parent---see the \rn{R-Go}~rule in Figure~\ref{fig:race.raw.waw.steps.go}.

\begin{figure}[!h]
  \centering
  \begin{ruleset}
    \input{rules/datarace/steps-go}
  \end{ruleset}
  \caption{Operational semantics augmented for race detection: thread creation\label{fig:race.raw.waw.steps.go}}
\end{figure}


%% file: rules/datarace/configs-simple.tex
\begin{equation}
  \label{eq:gomm.race.configs-simple}
  R \bnfdef \ngoroutine{p}{\idshb}{t}
  \bnfbar
  \ezwritesl{z}{v}{\idshbz{z}}
  \bnfbar 
  \emptyconf 
  \bnfbar
  R \parallel R\
  \bnfbar 
  \gchan{c}{q}
  \bnfbar 
  \nu n\  R\  .
\end{equation}


%% file: rules/datarace/steps-smem-raw-waw-simple.tex
\infrule{R-Write}{
  \idshbz{z} \subseteq \idshb 
  \andalso
  \fresh(m')
  \andalso
  \idshb'= \{ \ewritea{m'}{z} \} \union \idshb
  \andalso
  \idshbzp{z}= \{ \ewritea{m'}{z} \} \union \idshbz{z}
}{
  \ngoroutine{p}{\idshb}{\store{z}{v'}; t}
  \parallel
  \ezwritesl{z}{v}{\idshbz{z}}
  \trans{}
  \ngoroutine{p}{\idshb'}{t}
  \parallel
  \ezwritesl{z}{v'}{\idshbzp{z}}
}

\ruleskip

\infrule{R-Read}{
  \idshbz{z} \subseteq \idshb 
}{
  \ngoroutine{p}{\idshb}{\lets r=\ \load{z} \ins t}
  \parallel
  \ezwritesl{z}{v}{\idshbz{z}}
  \trans{}
  \ngoroutine{p}{\idshb}{\lets r=\ v \ins t}
  \parallel
  \ezwritesl{z}{v}{\idshbz{z}}
}


%% file: rules/datarace/steps-smem-excep-simple.tex
\infrule{R-Write-$\Texc_{WaW}$}{
  \idshbz{z} \not\subseteq \idshb
  \ifthenelse{\equal{\VERSION}{.full}}{
    \andalso
    \idshbz{z}\projectto{?} \subseteq \idshb
  }{}
}{
  \ngoroutine{p}{\idshb}{\store{z}{v'}; t}
  \parallel
  \ezwritesl{z}{v}{\idshbz{z}}
  \trans{}
  \TexcArg{ \idshbz{z} - \idshb }
}

\vspace{2pt}
\ruleskip

\ifthenelse{\equal{\VERSION}{.full}}{
\infrule{R-Write-$\Texc_{WaR}$}{
  \idshbz{z} \projectto{?} \nsubseteq \idshb
}{
  \ngoroutine{p}{\idshb}{\store{z}{v'}; t}
  \parallel
  \ezwritesl{z}{v}{\idshbz{z}}
  \trans{}
  \TexcArg{ \idshbr - \idshb }
}
\vspace{2pt}
\ruleskip
}{}

\infrule{R-Read-$\Texc_{RaW}$}{
  \idshbz{z} 
  \ifthenelse{\equal{\VERSION}{.full}}{
  \projectto{!}
  }{}
  \not\subseteq \idshb
}{
  \ngoroutine{p}{\idshb}{\lets r=\ \load{z} \ins t}
  \parallel
  \ifthenelse{\equal{\VERSION}{.full}}{
    \ezwritesl{z}{v}{\idshbz{z}}
  }{
    \ezwritesl{z}{v}{\idshbz{z}}
  }
  \trans{}
  \TexcArg{ \idshbz{z} - \idshb }
}


%% file: rules/datarace/steps-chans.tex
\infrule{R-Make}{
  q = [\lstateempty,\ldots,\lstateempty]
  \andalso
  \sizeof{q} = v
  \andalso
  \fresh(c)
}{
  \ngoroutine{p}{\idshb}{\lets r = \ \makechans{\T}{v} \ins t} 
  \widetrans{}
  \nuin{c}{(  \ngoroutine{p}{\idshb}{\lets r = c \ins t}   
  \parallel
  \gchanf{c}{}
  \parallel
  \gchanb{c}{q})
  }
}

\ruleskip

\infrule{R-Send}{
  \lnot \gisclosed (\gchanf{c}{q_2})
  \andalso
  \idshb' = \idshb+ \idshb''
}{
    \gchanb{c}{q_1::\idshb''} \parallel  
    \ngoroutine{p}{\idshb}{\gsend{c}{v};t} \parallel  
    \gchanf{c}{q_2}
    \widetrans{}
    \gchanb{c}{q_1} \parallel  
    \ngoroutine{p}{\idshb'}{t}
    \parallel \gchanf{c}{\esend{}{v}{\idshb}::q_2}
}

\ruleskip

\infrule{R-Rec}{
	v \not= \eot
  \andalso
  \idshb' = \idshb+ \idshb''
}{
  \begin{array}{rcl}
    \gchanb{c}{q_1} \parallel &
    \ngoroutine{p}{\idshb}{\lets r = \greceive{c}\ins t} &
    \parallel \gchanf{c}{q_2::\esend{}{v}{\idshb''}} \widetrans{} \\
    \gchanb{c}{\idshb::q_1} \parallel &
    \ngoroutine{p}{\idshb'}{\lets r = v \ins t} &
    \parallel \gchanf{c}{q_2}
  \end{array}
}

\ruleskip

\infrule{R-Rec$_\eot$}{
  \idshb' = \idshb + \idshb''
}{
  \ngoroutine{p}{\idshb}{\lets r = \greceive{c}\ins t}
  \parallel
  \gchanf{c}{\esend{}{\eot}{\idshb''}}
  \widetrans{}
  \ngoroutine{p}{\idshb'}{\lets r = \eot \ins t}
  \parallel
  \gchanf{c}{\esend{}{\eot}{\idshb''}}
}

\ruleskip

\infrule{R-Rend}{
  \idshb' = {\idshb}_1 + {\idshb}_2
}{
  \begin{array}{rcll}
	  \gchanb{c}{} \parallel &
    \ngoroutine{p_1}{{\idshb}_1}{\gsend{c}{v}; t} &
	  \parallel
    \ngoroutine{p_2}{{\idshb}_2}{\lets r = \greceive{c}\ins t_2} &
  	\parallel \gchanf{c}{} \widetrans{} \\

	  \gchanb{c}{} \parallel &
    \ngoroutine{p_1}{\idshb'}{t} &
	  \parallel
    \ngoroutine{p_2}{\idshb'}{\lets r = v\ins t_2} &
  	\parallel \gchanf{c}{}
  \end{array}
}

\ruleskip

\infrule{R-Close}{
	\lnot \gisclosed ( \gchanf{c}{q})
}{
  \ngoroutine{p}{\idshb}{\gcloses(c); t}
  \parallel
  \gchanf{c}{q}
  \widetrans{}
  \ngoroutine{p}{\idshb}{t}
  \parallel
  \gchanf{c}{\esend{}{\eot}{\idshb}:: q}
}
%

%% file: rules/datarace/steps-go.tex
\infrule{R-Go}{
  \fresh(p')
}{
  \ngoroutine{p}{\idshb}{\gos t'; t}
  \widetrans{}
  \nuin{p'}{(\ngoroutine{p'}{\idshb}{t'})}   \parallel \ngoroutine{p}{\idshb}{t}
}


%% file: race-full.tex
\subsection{Detecting write-after-read (WaR) races}
\label{sec:race.detector.war}

In the previous section, the detection of read-after-write and
write-after-write races required happens-before sets to contain write
labels only.  The detection of write-after-read races requires recording
read labels, as well.
A successful read of variable $z$ causes a fresh read label, say $m'$, to
be generated.  The pair $\ereada{m'}{z}$ is added to the reader's
happens-before set as well as to the record associated with $z$ in
memory---see rule~\rn{R-Read} of Figure~\ref{fig:race.steps.smem.full}.

\begin{figure}[!ht]
  \centering
  \begin{ruleset}
    \input{rules/datarace/steps-smem-simple}
    \ruleskip
  \end{ruleset}
  \caption{Operational semantics augmented for data-race detection\label{fig:race.steps.smem.full}}
\end{figure}

\begin{figure}[!ht]
  \centering
  \begin{ruleset}
    \input{rules/datarace/steps-smem-excep-simple}
    \ruleskip
  \end{ruleset}
  \caption{Exception conditions for WaR data-race detection\label{fig:race.steps.smem.excep.full}}
\end{figure}

In order for a write to memory to be successful, the writing thread must not only be aware of previous write events to a given shared variable, but must also account for all accumulated reads to the variable.  A write-after-read data-race is raised when a write is attempted by a thread and the thread is unaware of some previous reads to $z$.
In other words, there exist some read-label in the happens-before set associated with the variable's record, say $r \in \idshbz{z}\projectto{?}$, that is not in the thread's happen-before set, $r \notin \idshb$.
The projection $\projectto{?}$ essentially filters out write events from the happens-before set.
Under these circumstances, the precondition $\idshbz{z}\projectto{?} \nsubseteq \idshb$ of the \rn{R-Write-$\Texc_{WaR}$}~rule is met and a race is reported.

Compared to the detector of Section~\ref{sec:race.detector.raw.waw}, the
reporting of WaW races in rule~\rn{R-Write-$\Texc_{WaW}$} is augmented with
the precondition $\idshbz{z}\projectto{?} \subseteq \idshb$.  Without this
precondition, there would be non-determinism when reporting WaW and WaR
conflicts.%
\footnote{Consider the scenario in which $p$ writes to and then reads from
  the shared variable $z$.  Say the write to $z$ generates a label $w$ and
  the read generates $r$.  If a thread $p'$ attempts to write to $z$
  without first communicating with $p$, $p'$ will not be aware of the prior
  read and write events.  In other words, the happens-before set of $p'$
  will contain neither $\ewritea{w}{z}$ nor $\ereada{r}{z}$.  Both
  rules~\rn{R-Write-$\Texc_{WaW}$} and~\rn{R-Write-$\Texc_{WaR}$} are
  enabled in this case.  However, the read happens-after the write that
  generated $\ewritea{w}{z}$.}  Note, however, that when both WaW and WaR
apply, the read in the WaR race happens-after the write involved in the WaW
race.  We favor to resolve this non-determinism and to report the most
recent conflict.

The detector presented here can flag all conflicts: read-after-write, write-after-write, and write-after-read.
In Section~\ref{sec:race.detector.gc} we also make the detector efficient by ``garbage collecting'' stale information.  But before then, let us look at a couple of examples that illustrate the detector's operation.


%% file: rules/datarace/steps-smem-simple.tex
\infrule{R-Write}{
  \idshbz{z} \subseteq \idshb
  \andalso
  \fresh(m')
  \andalso
  \idshb'= \{ \ewritea{m'}{z} \} \union \idshb
  \andalso
  \idshbzp{z}= \{ \ewritea{m'}{z} \} \union \idshbz{z}
}{
  \ngoroutine{p}{\idshb}{\store{z}{v'}; t}
  \parallel
  \ezwritesl{z}{v}{\idshbz{z}}
  \trans{}
  \ngoroutine{p}{\idshb'}{t}
  \parallel
  \ezwritesl{z}{v'}{\idshbzp{z}}
}

\vspace{6pt}
\ruleskip

\infrule{R-Read}{
  \idshbz{z}\projectto{!} \subseteq \idshb
  \andalso
  \fresh(m')
  \andalso
  \idshb'= \{ \ereada{m'}{z} \} \union 
  \idshb
  \andalso
  \idshbrp= \{ \ereada{m'}{z} \} \union 
  \idshbr
}{
  \ngoroutine{p}{\idshb}{\lets r=\ \load{z} \ins t}
  \parallel
  \ezwritesl{z}{v}{\idshbz{z}}
  \trans{}
  \ngoroutine{p}{\idshb'}{\lets r=\ v \ins t}
  \parallel
  \ezwritesl{z}{v}{\idshbzp{z}}
}


%% file: race-gc.tex
\section{Efficient data-race detection}
\label{sec:race.detector.gc}
\def\VERSION{.gc}

We have been gradually introducing a data-race checker.  In Section~\ref{sec:race.detector.raw.waw}, we presented a simple checker that flags after-write races (WaW and RaW) but is not equipped for write-after-read (WaR) detection.  In Section~\ref{sec:race.detector.war}, we augmented the detector to handle WaR.  Here, we discuss how these detectors can be implemented efficiently; where efficiency is gained by employing ``garbage collection'' to reduce the detector's memory footprint.
Note that keeping one record per variable is already a form of efficiency gain.  In a relaxed memory model, since there may be more than one value associated with a variable at any point in the execution, one might keep one record per memory event~\cite{fava*:oswmm-chan-journal}.  The first step towards a smaller footprint is to realize that, if the underlying memory model supports the DRF-SC guarantee, a data-race detector can be built assuming sequential consistency.
The reason being that, when a data race is flagged, execution stops at the point in which the weak and strong memory models' executions would diverge.

Knowing that memory events can overtake each other, in this section we discuss how stale or redundant information can be garbage collected.  More precisely, we show how to garbage collect the data structures that hold happens-before information, that is, the thread-local happens-before set and the per-memory-location one.

\subsection{Most recent write}
\label{sec:race.detector.gc.msr}

Terms representing a memory location have taken different shapes when compared to the undecorated semantics.  In the undecorated semantics, the content $v$ of a variable $z$ is written as a pair $\ewritesl{}{z}{v}$.
For after-write race detection, an entry in memory took the form of $\ezwritesl{z}{v}{\idshbz{}}$ with $\idshbz{}$ holding information about prior write events.
Our first optimization comes from realizing that we do not need to keep a set of prior write events.  We can record only the most recent write and still be able to flag all after-write racy \emph{executions.}
With this optimization, we may fail to report all \emph{accesses} involved in the race, but we will still be able to report the execution as racy and to flag the most recent conflicting write event.
This optimization is significant; it reduces the arbitrarily large set of prior write events to a single point.

%
An intuitive argument for the correctness of the optimization comes from noticing that a successful write to a variable can be interpreted as the writing thread taking \emph{ownership} of the variable.
Suppose a goroutine $p$ has just updated variable $z$.  At this point, $p$ is only goroutine whose happens-before set contains the label, say $m$, associated with this write-record.
The placement of the new label into $p$'s happens-before set can be seen as recording $p$'s \emph{ownership} of the variable: a data-race is flagged if any other thread attempts to read or write to $z$ without first synchronizing with $p$---see the check $\ewritea{m}{z} \in \idshb$ in the premise of the \rn{R-Write} and \rn{R-Read} rules of Figure~\ref{fig:race.steps.smem.gc}.

When $p$ sends a message onto a channel, the information about $m$ is also sent.  Suppose now that a thread $p'$ reads from the channel and receives the corresponding message before $p$ makes any further modifications to $z$.
The tuple $\ewritea{m}{z}$ containing the write-record's label is added to $p'$'s happens-before set.  Now both $p$ and $p'$ are aware of $z$'s most recent write to $z$.
The existence of $m$ in both goroutine's happens-before sets imply that either $p$ or $p'$ are allowed to update $z$'s value.  We can think of the two goroutines as \emph{sharing} $z$.  Among $p$ and $p'$, whoever updates $z$ first (re)gains the \emph{exclusive} rights to $z$.

It may be worth making a parallel with hardware and cache coherence protocols.  Given the derivation rules, we can write a race detector as a state machine. \index{MESI}Compared to the Modified-Exclusive-Shared-Invalid protocol (MESI), our semantics does not have the \emph{modified} state: all changes to a variable are immediately reflected in the configuration, there is no memory hierarchy in the memory model.
As hinted above, the other states can be interpreted as follows: If the label of the most recent write to a variable is only recorded in one goroutine's happens-before set, then we can think of the goroutine as having \emph{exclusive} rights to the variable.
When a number of goroutines contain the pair $\ewritea{m}{z}$ in their happen-before set with $m$ being the label of the most recent write, then these goroutines can be thought to be \emph{sharing} the variable.  Other goroutines that are unaware of the most recent write can be said to hold \emph{invalid} data.

\subsection{Runtime configuration and memory related reduction rules}
\label{sec:race.detector.gc.rc}

Given the ``most recent write'' optimization above, and, if we were satisfied with after-write conflicts, an entry in memory would take the form of $\ewritesl{m}{z}{v}$, with the label $m$ uniquely identifying the event associated with $v$ having been stored into $z$.
Being able to flag after-write but not write-after-read races may be an adequate trade-off between completeness and efficiency.  By not having to record read events, a simplified detector tailored for after-write race detection has a much smaller footprint than when read-after-write conflicts are also taken into account.  Besides, a write-after-read race that is not flagged in an execution may realize itself as a read-after-write race in another run, and then be flagged by the simplified detector.%
\footnote{Intuitively, say $S_0 \trans{e_0} S_1 \trans{e_1} \cdots \trans{e_{n-1}} S_n$ is a run starting from an initial configuration $S_0$.
Let $\ind$ be an independence relation on events, meaning, 
given $S_i \trans{e_i} S_{i+1} \trans{e_{i+1}} S_{i+2}$,
we say that $e_i \ind e_{i+1}$ if there exist $S'$ such that
$S_i \trans{e_{i+1}} S' \trans{e_i} S_{i+2}$.
The independence relation induces an equivalence relation on traces, namely, traces are equivalent if they can be derived from one another via the permutation of independent events.
It can be shown that if $S_0 \trans{h} S_n$ is a run containing a write-after-read race, the exist an equivalent run in which the race materializes as a read-after-write race.}

In contrast, the detection of write-after-read races requires more book-keeping: we need read- in addition to write-labels.  This addition is required because a WaR conflict can ensue between an attempted write and \emph{any} previous unsynchronized read to the same variable.
Therefore, the race-checker is made to remember all such potentially troublesome reads.\footnote{Since depending on scheduling, a WaR data-race can manifest itself as RaW race, one option would be not add instrumentation for WaR race detection and, instead, hope to flag the RaW manifestation instead.  Such practical consideration illustrates the trade-off between completeness versus run-time overhead.}
The runtime configuration is thus modified, this time as to contain entries of the form $\erwritesl{m}{z}{v}{\idshbr}$.  The label $m$ identifies of the most recent write event to $z$ and the set $\idshbr$ holds-read event identifiers, namely, the identifiers of reads that accumulated after $m$.
\input{./rules/datarace/configs}
%
\noindent
Note that \emph{records} of the form $\erwritesl{m}{z}{v}{\idshbr}$ can be seen as $n+1$ recorded events: one write together with $n \geq 0$ read events.  \ifreport Writing the recorded read events as in equation~(\ref{eq:race.parallelreadevents}), the notation $\erwritesl{m}{z}{v}{\idshbr}$ is a compact short-hand for $\ewritesl{m}{z}{v_1} \parallel \ereadfrom{r_1}{m} \parallel \ereadfrom{r_2}{m} \parallel \ldots$. \fi

The formal semantics maintains the following invariants.  First, the happens-before information $\idshbr$ in $\erwritesl{m}{z}{v}{\idshbr}$ contains information of the form $\ereada{m'}{z}$ only, i.e., there are no write events and all read-events concern variable $z$. 
Also, the event labels are unique for both reads and writes.
In an abuse of notation, we may refer to $m$ being in $\idshbr$ and write $m \in \idshbr$ meaning, more precisely, $\ereada{m}{z} \in \idshbr$.

\begin{figure}[!ht]
  \centering
  \begin{ruleset}
    \input{rules/datarace/steps-smem-gc}
    \ruleskip
  \end{ruleset}
  \caption{Operational semantics augmented for efficient data-race detection\label{fig:race.steps.smem.gc}}
\end{figure}

\subsection{Garbage collection of happens-before sets}
\label{sec:race.detector.gc.ehb}

Knowledge of past events contained in a happens-before set $\idshb$ is naturally monotonically increasing.  For example, each time a goroutine learns about happens-before information, it adds to its pool of knowledge. In particular, events that are known to have ``happened-before'' cannot, by learning new information, become ``concurrent.''
An efficient semantics, however, does not accumulate happens-before information indiscriminately; instead, it purges redundant information.  We say ``redundant'' for the purpose of flagging racy executions, but leaving out conflicting accesses that have been overtaken by more recent memory events.

\subsubsection{Garbage collection on \emph{writes}}
\label{sec:race.detector.gc.ehb.writes}

For a thread $t$ to successfully write to $z$, all previously occurring accesses to $z$ must be in happens-before with the thread's current state.
One optimization comes from realizing that we can purge all information about prior accesses the variable $z$ from the happens-before set of the writing thread $t$.
We call these prior accesses \emph{redundant} from the point of view of flagging racy executions.
The reason for the correctness of this optimization is as follows:
All future access of $t$ to $z$ are synchronized with the redundant accesses, after all, the accesses are recorded in $t$'s happens-before set.  Therefore, from the perspective of $t$, these accesses do not affect data-race detection.
For the same reason, if a thread $t'$ synchronizes with $t$, there is no race to report if and when $t'$ accesses memory---the absence of these redundant accesses from $t'$'s happens-before is, therefore, inconsequential.
Finally, if $t'$ does not synchronize with $t$, then an access to $z$ is racy because it is unsynchronized with $t$'s most recent write, regardless of the redundant prior accesses.
Note that this optimization allows us to flag all racy \emph{executions} even if we fail to report some of the accesses involved in the race.

Rule~\rn{R-Write} of Figure~\ref{fig:race.steps.smem.gc} embodies this discussion.
Before writing, the rule checks that the attempted write happens-after all previously occurring accesses to $z$.
This check is done by two premises: premise $\ewritea{m}{z} \in \idshb$ makes sure that the most recent write to $z$, namely, the one that produced event $\ewritea{m}{z}$, is in happens-before with the current thread state $\idshb$.  As per discussion in Section~\ref{sec:race.detector.gc.msr}, being synchronized with the most recent write means the thread is synchronized with all writes up to that point in the execution.  The other premise, $\idshbr \subseteq \idshb$, makes sure that the attempted write is in happens-after read accesses to $z$.
If these two premises are satisfied, the write can proceed and prior accesses to $z$ are garbage collected from the point of view of $t$.  The filtering of redundant accesses is done by subtracting $\idshb \projectto{z}$ in
\[ 
  \idshb'= \{ \ewritea{m'}{z} \} \union
          \left( \idshb - \idshb \projectto{z}
          \right)
\]
where $\projectto{z}$ projects the happens-before set down to operations on variable $z$.
Finally, the write rule also garbage collects the in-memory record $\idshbr$ by setting it to $\emptyset$,%
\footnote{As per discussion in Section~\ref{sec:race.detector.gc.msr}, a term representing a memory location $\erwritesl{m}{z}{v}{\idshbr}$ records in $\idshbr$ all the reads to $z$ that have accumulated after the write  that generated the write label $m$.  When a new write $m'$ of value $z:=v'$ ensues, we update the memory term to record this new write and we reset its corresponding $\idshbr$ to $\emptyset$.} meaning that no read event have accumulated after the write yet.

\subsubsection{Garbage collection on \emph{reads}}
\label{sec:race.detector.gc.ehb.reads}

We also garbage collect on load operations.
Say $t$ reads from $z$, thus generating event $\ereada{m'}{z}$.
Let us call \emph{redundant} the memory accesses to $z$ in $t$'s happens-before set at the time event $\ereada{m'}{z}$ takes place, with the exception of $\ewritea{m}{z}$.
A read operation can only conflict with a future write; there are not read-read conflicts.  For a future write to take place, the writing thread will need to synchronize with a thread that ``knows'' about the read $m'$.%
\footnote{``Knowing about the read $m'$'' is a necessary condition for a thread to successfully write to $z$, but it is not a sufficient one.  There may exist other reads, say $m''$, $m'''$, etc that are concurrent with $m'$.  A thread needs to synchronize with all such concurrent reads before it can successfully write to $z$.}
Any thread that knows of $m'$ would also know about the redundant access to $z$ and know of $\ewritea{m}{z}$.  In other words, $m'$ and $m$ subsume all happened-before accesses of $z$ from the perspective of $t$.
Therefore, we can garbage collect all such accesses by filtering them out of the thread's happen-before set, as in
\[
  \idshb'= \{ \ereada{m'}{z} \} \union
    \left( \idshb - \idshb \projectto{z} \right)
    \union \{ \ewritea{m}{z} \}
    \text{.}
\]
These redundant accesses are also filtered out of the in-memory happens-before set:
\[
  \idshbrp= \{ \ereada{m'}{z} \} \union
    \left( \idshbr - \idshb \projectto{z} \right)
    \text{.}
\]

\subsubsection{Off-line garbage collection and channel communication}
\label{sec:race.detector.gc.ehb.offline}

The garbage collector rules of Figure~\ref{fig:race.steps.gc} can be run non-deterministically during the execution of a program.  
Rule \rn{R-GC} eliminates stale entries from the happens-before set of a thread.
It can be sensible to perform garbage collection also \emph{after} a thread interacts with a channel, as happens-before information communicated via channels are likely to become stale.
For example, suppose a thread, whose happens-before set does not contain stale entries, sends onto a channel and continues executing.  By the time a receive takes place, the happens-before set transmitted via the channel may have become stale.  Similarly for happens-before transmitted between receives and prior sends via the backward channel.
Alternatively, we may choose an implementation in which the happens-before of inflight messages are also gargabe collected, in which case we would process the happens-before sets in a channel's forward and backquard queues.

\begin{figure}[!ht]
  \centering
  \begin{ruleset}
    \input{rules/datarace/steps-gc}
    \ruleskip
  \end{ruleset}
  \caption{Off-line garbage collection\label{fig:race.steps.gc}}
\end{figure}


%% file: rules/datarace/configs.tex
\begin{equation}
  \label{eq:gomm.race\VERSION.configs}
  R \bnfdef \ngoroutine{p}{\idshb}{t}
  \bnfbar
  \erwritesl{m}{z}{v}{\idshbr}
  \bnfbar 
  \emptyconf 
  \bnfbar
  R \parallel R\
  \bnfbar 
  \gchan{c}{q}
  \bnfbar 
  \nu n\  R\  .
\end{equation}


%% file: rules/datarace/steps-smem-gc.tex
\infrule{R-Write}{
  \ewritea{m}{z}  \in \idshb 
  \andalso 
  \idshbr \subseteq \idshb
  \andalso
  \fresh(m')
  \andalso
  \idshb'= \{ \ewritea{m'}{z} \} \union 
          \left( \idshb 
                 - \idshb \projectto{z}
          \right)
}{
  \ngoroutine{p}{\idshb}{\store{z}{v'}; t}
  \parallel
  \erwritesl{m}{z}{v}{\idshbr}
  \trans{}
  \ngoroutine{p}{\idshb'}{t}
  \parallel
  \erwritesl{m'}{z}{v'}{\emptyset}
}

\vspace{10pt}
\ruleskip

\infrule{R-Read}{
  \begin{array}{lcr}
  &
  &
  \idshbrp= \{ \ereada{m'}{z} \} \union
    \left( \idshbr - \idshb \projectto{z} \right)
    \phantom{\{ \union \ewritea{m}{z} \}}
  \\
  \ewritea{m}{z} \in \idshb
  \andalso
  & 
  \fresh(m')
  \andalso
  &
  \idshb'= \{ \ereada{m'}{z} \} \union
    \left( \idshb - \idshb \projectto{z} \right)
    \union \{ \ewritea{m}{z} \}
  \end{array}
}{
  \ngoroutine{p}{\idshb}{\lets r=\ \load{z} \ins t}
  \parallel
  \erwritesl{m}{z}{v}{\idshbr}
  \trans{}
  \ngoroutine{p}{\idshb'}{\lets r=\ v \ins t}
  \parallel
  \erwritesl{m}{z}{v}{\idshbrp}
}


%% file: rules/datarace/steps-gc.tex
\infrule{R-GC}{
  \begin{array}{rl}
    \idshb'= \idshb
    & 
    -\ \{
      \ewritea{\hat{m}}{z} \suchthat 
      \ewritea{\hat{m}}{z} \in \idshb
      ~\land~
      \hat{m} \neq m
    \}
    \\
    &
    -\ \{
      \ereada{\hat{m}}{z} \suchthat 
      \ereada{\hat{m}}{z} \in \idshb
      ~\land~
      \ereada{\hat{m}}{z} \notin \idshbr
    \}
  \end{array}
}{
  \ngoroutine{p}{\idshb}{t}
  \parallel
  \erwritesl{m}{z}{v}{\idshbr}
  \trans{}
  \ngoroutine{p}{\idshb'}{t}
  \parallel
  \erwritesl{m}{z}{v}{\idshbr}
}


%% file: comparison.tex
\section{Comparison with vector-clock based race detection}
\label{sec:race.comparison}

Vector clocks (VCs) are a mechanism for capturing the happen-before relation over events emanating from a program's execution~\cite{mattern:virtual}.  A \emph{vector clock} $\vc{V}$ is a function $\kw{Tid} \rightarrow \kw{Nat}$ which records a clock, represented by a natural number, for each thread in the system.
``VCs are partially-ordered ($\vcleq$) in a pointwise manner, with an associated join operation ($\vcjoin$) and minimal element ($\vcbot{V}$).  In addition, the helper function $\vcinc{t}$ increments the $t$-component of a VC''~\cite{flanagan.freund:fasttrack}.
\begin{align*}
  \vc{V}_1 \vcleq \vc{V}_2 ~~&\text{iff}~~ \forall t.~ \vc{V}_1(t) \leq \vc{V}_2(t) \\
  \vc{V}_1 \vcjoin \vc{V}_2 ~~&=~~ \lambda t.~ \maxs(\vc{V}_1(t), \vc{V}_2(t)) \\
  \vcbot{V}~~&=~~ \lambda t.~ 0\\
  \vcinc{t}(\vc{V}) ~~&=~~ \lambda u.~ \ifs u=t \thens \vc{V}(u)+1 \elses \vc{V}(u)
\end{align*}

Using vector clocks, \citet{pozniansky.schuster:efficient} proposed a data-race detection algorithm referred to as \djitp\
Their algorithm works as follows.
Each thread $t$ is associated with a vector clock $\vc{C}_t$.  Entry $\vc{C}_t(t)$ stores the current time at $t$, while $\vc{C}_t(u)$ for $u \neq t$ keeps track of the time of the last operation ``known'' to $t$ as having been performed by thread $u$.

The algorithm also keeps track of memory operations.  Each memory location $x$ has two vector clocks, one associated with reads, $\vc{R}_x$, and another with writes, $\vc{W}_x$.  
The clock of he last read from variable $x$ by thread $t$ is recorded in $\vc{R}_x(t)$; similar for $\vc{W}_x(t)$ and writes to $x$ by $t$.
When it comes to reading from memory, a race is flagged when a thread $t$ attempts to read from $x$ while being ``unaware'' of some write to $x$.  Precisely, a race is flagged when $t$ attempts to read from $x$ and there exists a write to $x$ by thread $u$, $\vc{W}_x(u)$, that is not accounted for by $t$, meaning $\vc{W}_x(u) \geq \vc{C}_t(u)$, or, equivalently, $\vc{W}_x \not\vcleq \vc{C}_t$.
If $t$ succeeds in reading from $x$, then $\vc{R}_x(t)$ is updated to the value of $\vc{C}_t(t)$.
When it comes to writing to memory, a race is flagged when $t$ attempts to write to $x$ while being unaware of some read or write to $x$, meaning either $\vc{R}_x \not\vcleq \vc{C}_t$ or $\vc{W}_x \not\vcleq \vc{C}_t$.
If $t$ succeeds in writing to $x$, then $\vc{W}_x(t)$ is updated to $\vc{C}_t(t)$.

A thread's clock is advanced when the thread executes synchronization operations, which have bearing on the happens-before relation.  The algorithm was proposed in the setting of locks; each lock $m$ is associated with a vector clock $\vc{L}_m$.  
When a thread $t$ acquires $m$, then $\vc{C}_t$ is updated to $\vc{C}_t \vcjoin \vc{L}_m$.
Acquiring a lock is analogous to receiving from a channel with buffer size one: the receiving thread updates its vector clock by incorporating the VC previously ``stored'' in the lock.
When a thread $t$ releases a lock $m$, the vector clock $\vc{L}_m$ is updated to $\vc{C}_t$ and thread's clock is advanced, meaning $\vc{C}_t := \vcinc{t}(\vc{C}_t)$.
We can think of lock release as placing a message, namely the vector clock associated with the releasing thread, into a buffer of size one.
Thus, in comparison with the approach presented in our paper, lock operations are a special case of buffered channel communication.
Our paper deals with channels of arbitrary size and their capacity limitations.

A significant difference between our approach and \djit\ is that we dispense with the notion of vector clocks.  Vector clocks are a conceptual vehicle to capturing partial order of events.  Instead of relying on VCs, our formalization is tied directly to the concept of happens-before.
Vector clocks are expensive.
VCs require $O(\nthreads)$ storage space and common operations on VCs consume $O(\nthreads)$ time where $\nthreads$ is the number of entries in the vector.  In the case of race detection, $\nthreads$ is the number of threads spawn during the execution of a program.
It turns out that not all uses of VCs in \djit\ are strictly necessary.
In fact, \citet{flanagan.freund:fasttrack} introduce the concept of \emph{epoch}, which consists of a pair $\epoch{c}{t}$ where $c$ is a clock and $t$ a thread identifier.
They then replace $\vc{W}_x$, the vector clock tracking writes to $x$, with a single epoch.  This epoch captures the clock and thread identity associated with the most recent write to $x$.
Similarly, in our approach, a memory location is associated with the identifier of only the most-recent write to that location.  
Any thread who is ``aware'' of this identifier is allowed to read from the corresponding variable.

\fasttrack\ also reduces the dependency on vector clocks by replacing $\vc{R}_x$ with the epoch of the most recent read to $x$.
However, since reads are not totally ordered, \fasttrack\ dynamically switches back to a vector clock representation when needed.
Similar to \fasttrack, we record the most recent (unordered) reads which, in the best case, involves an $O(1)$-memory footprint and $O(\nthreads)$ at the worst.

When it comes to per-thread memory consumption, however, our approaches look very different.
While \djit's and \fasttrack's worst-case memory consumption per thread is $O(\nthreads)$, our is $O(\nvars \nthreads)$ where $\nvars$ is the number of shared variables in a program.%
\footnote{We believe the worst case is a degenerate case unlikely to happen: it involves every thread reading from every shared variable and then exchanging messages as to inform everyone else about their read events.}
Vector clocks' memory efficiency, when compared to happens-before sets, come from VC's ability to succinctly capture the per-thread accesses that take place in between advances of a clock.
A thread's clock is advanced when the thread releases a lock.%
\footnote{If channels were used instead of locks, the advance would take place when a thread sends onto or receives from a channel.}
All accesses made by a thread $t$ in a given clock $c$ are captured by the clock: if another thread $u$ ``knows'' the value $c$ of $t$'s clock, then $u$ is in happens-after with \emph{all} accesses made by $t$---that is, all accesses up to when $t$'s clock was advanced to $c+1$.
In contrast, the happens-before set representation is much more coarse.  We keep track of individual accesses, as opposed to lumping them together into a clock number.  This coarseness explains the extra factor of $\nvars$ in the worst-case analysis of the happens-before set solution.
Although being a disadvantage in the worst case scenario, it does provide benefits, as we discuss next.

Note that the vector-clocks associated with threads and locks grow monotonically.
By \emph{growing monotonically} we do not mean that time marches forward to increasing clock values.  
Instead, we mean that the number of clocks in a vector grows without provisions for the removal of entries.
This growth can lead to the accumulation of ``stale'' information, where by stale we mean information that is not useful from the point of view of race detection.
This growth stands in contrast to our approach to garbage collection.
Stale information is purged from happens-before sets, which means they can shrink back to size zero after having grown in size.

Let us look at an example that illustrates this difference in treatment of stale information.  Consider the producer/consumer paradigm, where a thread produces information to be consumed by other threads.
Say $p_0$ produces information by writing to the shared variable $z$.
The thread then notifies consumers, $p_1$ and $p_2$, by sending a message on channel $c$.
The consumers read from $z$ and signal the fact that they are done consuming by sending onto channel $d$.
The producer $p_0$ writes to $z$ again once it has received the consumers' messages.

\begin{equation*}
\begin{array}{lccll}
\text{Producer} & && \multicolumn{2}{c}{\text{Consumers}} \\
~~~~~ p_0 & && ~~~ p_1 & ~~~ p_2 \\
\hline \\[-4pt]
\store{z}{42}; & \qquad\qquad & \qquad\qquad &
  \greceive{c}; \qquad\qquad\qquad\qquad &
  \greceive{c}; \\
\gsend{c}{0};\ \gsend{c}{0}; & &&
  \load{z}; &
  \load{z}; \\
\greceive{d};\ \greceive{d}; & &&
  \gsend{d}{0} & 
  \gsend{d}{0} \\
\store{z}{43} & && &
\end{array}
\end{equation*}

Let us run this example against a prototype implementation \cite{www:grace} of our proposed race detector, called \grace, and against \fasttrack. 
Consider the point in the execution after $p_0$ has written to $z$, the consumers have read from $z$ and notified $p_0$, and $p_0$ is about to write to $z$ again.
Below is the state of the detectors at this point.  The information contained in the happens-before sets and the vector-clocks is very similar.  There are three entries for $p_0$, and two entries for $p_1$ and $p_2$ each.

\bigskip

\begin{tabular}{lcl}
Happens-before sets &  \qquad\qquad\qquad & Vector-clocks\\
\hline  \\[-4pt]
$ \idshb^{p_0} = \{ \ewritea{m_0}{z}, \ereada{m_1}{z}, \ereada{m_2}{z} \}$ 
  && $\vc{C}_{p_0} = \vcbot{}[ p_0 \mapsto 6, p_1 \mapsto 1, p_2 \mapsto 1 ]$ \\[4pt]
$ \idshb^{p_1} = \{ \ewritea{m_0}{z}, \ereada{m_1}{z} \}$ 
  && $\vc{C}_{p_1} = \vcbot{}[ p_0 \mapsto 2, p_1 \mapsto 2 ]$ \\[4pt]
$ \idshb^{p_2} = \{ \ewritea{m_0}{z}, \ereada{m_2}{z} \}$
  && $\vc{C}_{p_2} = \vcbot{}[ p_0 \mapsto 3, p_2 \mapsto 2 ]$
\end{tabular}

\bigskip

The happens-before set $\idshb^{p_0}$ show the reads by $p_1$ and $p_2$ as being in happens-before with respect to $p_0$, along with $p_0$'s own write to $z$.
It also shows $p_1$ and $p_2$ as being ``aware'' of $p_0$'s write to $z$, as well as being ``aware'' of their own reads to $z$.
The same information is captured by the vector clocks.
Recall that the bottom clock, $\vcbot{}$, maps every process-id to the clock value of $0$.  Thus, the VC associated with $p_0$ contains $p_0$'s clock (which happens to be 6) as well as the clock associated with the reads to $z$ by $p_1$ and $p_2$.
In this execution, $p_0$'s clock was $2$ when the thread wrote to $z$.
Thus, the entry $p_0 \mapsto 2$ in $\vc{C}_{p_1}$ and the entry $p_0 \mapsto 3$ in $\vc{C}_{p_2}$ place the write to $z$ by $p_0$ in $p_1$'s and $p_2$'s past.

The difference between our approach the VC based approach is evidenced in the next step of execution, when $p_0$ writes to $z$ for the second time.
This write subsumes all previous memory interactions on $z$.  In other words, this write is in happens-after with respect to all reads and writes to $z$ up to this point in the execution of the program.  Therefore, it is sufficient for a thread to synchronize with $p_0$ before issuing a new read or write to $z$; also, it is no longer necessary to remember the original write to $z$ and the reads from $z$ by $p_1$ and $p_2$.
Here are the happens-before sets and vector-clocks in the next step of execution, meaning, after $p_0$ writes to $z$ the second time:

\bigskip

\begin{tabular}{lcl}
Happens-before sets &  \qquad\qquad\qquad & Vector-clocks\\
\hline  \\[-4pt]
$\idshb^{p_0} = \{ \ewritea{m_3}{z} \} \phantom{, \ereada{m_1}{z}, \ereada{m_2}{z}}$ 
  && $\vc{C}_{p_0} = \vcbot{}[ p_0 \mapsto 6, p_1 \mapsto 1, p_2 \mapsto 1 ]$ \\[4pt]
$\idshb^{p_1} = \{ \}$
  && $\vc{C}_{p_1} = \vcbot{}[ p_0 \mapsto 2, p_1 \mapsto 2 ]$ \\[4pt]
$\idshb^{p_0} = \{ \}$ 
  && $\vc{C}_{p_2} = \vcbot{}[ p_0 \mapsto 3, p_2 \mapsto 2]$ \\[4pt]
\end{tabular}

\bigskip

The happens-before sets are mostly empty; the only entry corresponds to the most recent write to $z$, which is known to $p_0$.
Meanwhile, the vector clocks are unchanged.
Note, however, that every entry with the exception of $p_0 \mapsto 6$ in $\vc{C}_{p_0}$ is stale.
In other words, with the exception of $p_0 \mapsto 6$, the presence or absence of all other entries does not alter a thread's behavior.
To illustrate this point, take entry $p_0 \mapsto 2$ in $\vc{C}_{p_1}$ as an example: if $p_1$ were to attempt to access $z$, a data race will ensue regardless of whether or not the entry $p_0 \mapsto 2$ is in $p_1$'s vector clock.
Therefore, ideally, we would want these stale entries purged from the vector-clocks of $p_0$, $p_1$, and $p_2$.  Concretely, we would want $\vc{C}_{p_0} = \vcbot{}[ p_0 \mapsto 6 ]$ and $\vc{C}_{p_1} = \vc{C}_{p_2} = \vcbot{}$.

Similar unbounded growth occurs in the VCs associated with locks,\footnote{The \emph{acquire} grows the VC associated with the acquiring thread; the \emph{release} sets the VC of the corresponding lock to the VC of the acquiring thread.} thus also leading to the accumulation of stale information.
We conjecture that an approach that purges stale information from VCs, similar to our notion of garbage collection, would be highly be beneficial.
VC-based implementations are very efficient in managing the memory overhead associated with variables.  For example TSan, a popular race-detection library based on vector clocks and that comes with the Go tool chain, stores one write and a small number of reads per memory location (the number of reads stored is 4 in the current implementation) \cite{google.threadsanitizer}.  Capping the number of tracked read events leads to false negatives; the cap a fair compromise between \emph{recall} and memory consumption.
In order to further reduce the memory foot-print of modern race detection implementations, we are thus left with devising approaches to managing threads' and locks' memory overhead.  

Unfortunately, reducing memory pressure on vector-clocks associated with threads and locks is arguably more difficult than reducing memory pressure on VCs associated with shared variables.
In one hand, if a variable does not ``remember'' a read or write to itself as having happened-before, then the variable becomes more permissive from the point of view of race detection; meaning, more threads would be able to interact with this variable without raising a data-race, even when races should have been reported.
On the other hand, if a \emph{thread} ``forgets'' about some prior read or write access that have taken place on a variable, a spurious data race may be raised.
Thus, while dropping clock entries in the VCs associated with variables can introduce false negatives, dropping clock entries from VCs associated with threads and locks introduce false positives.
From a practical perspective, false negatives are acceptable and can even be mitigated,%
\footnote{Provided we run a program enough times, we can randomly evict entries from a VC or happens-before set associated with a variable such that we eventually flag all existing races of the program.}
however, being warned of non-existing races is overwhelming to the application programmer, which means false positives are generally not tolerated.

\iftiming
\input{comparison_timing}
\fi

%% file: discussion.tex
\section{Connections with trace theory}
\label{sec:race.discussion}

Our operational semantics mimics the Go memory model in defining synchronization in terms of channel communication.  Specifically, we abide by rules~(\ref{eq:gomm.hb.chan.forward}) and~(\ref{eq:gomm.hb.chan.backward}), which establish a happens-before relation between a send and the completion of its corresponding receive, and, due to the boundedness of channels, between a receive and the completion of a future send.
However, these are not the only imposition by the semantics on the order of events.  Channels act as FIFO queues in both Go~\cite{www:golang-spec-channel-types} as well as in our operational semantics. However, neither Go nor our operational semantics establish a happens-before relation between consecutive sends or consecutive receives.
For example, the $i^{\mathit{th}}$ send on a channel $c$ does not happens-before the $(i+1)^{\mathit{th}}$ send on $c$.
Therefore, there exist events that are necessarily ordered, but that are not in happens-before relation.  

It is tempting to think of happens-before in terms of observations, where $a$ and $b$ are in happens-before if and only if we observe $a$ followed by $b$, and never the other way around.
This intuition is captured by the following tentative definition:

\vspace{10pt}
\noindent
\textit{Let $\idx{a}{h}$ be the index of event $a$ in a run $h$.}
\textit{Given the set of runs $H$ starting from an initial configuration,
we say that event $a$ \emph{happens-before} $b$ 
if-and-only-if, for all runs $h \in H$ such that $a, b \in h$,
$\idx{a}{h} < \idx{b}{h}$.}
\vspace{10pt}

When it comes to weak memory systems, there exist events that are ordered according to the above tentative definition but that are not in happens-before relation.
Take the improperly synchronized message-passing example of Figure~\ref{fig:mp.prog} as an example.
In this example, a thread $p_0$ writes to a shared variable \verb|z| and sets a flag; another thread, $p_1$, checks the flag reads from \verb|z| if the flag has been set.
\begin{figure}[hbt]
  \centering
  \begin{tabular}{|lc|lc|}
    \hline
    \multicolumn{2}{|c|}{$p_0$} &
    \multicolumn{2}{|c|}{$p_1$} \\
    \hline
    ~\texttt{z~~~ := 42;}~~ & $(A)$~~~~~~~~ & ~\texttt{r = load done;}~~~~ & $(C)$~~ \\
    ~\texttt{done := true;}~~ & $(B)$~~~~~~~~ & ~\texttt{if r then}~~~~ & \\
                            & & ~\qquad\texttt{load z}~~~~ & $(D)$~~ \\
      \hline
    \end{tabular}
  \caption{Message passing example.\label{fig:mp.prog}}
\end{figure}

If $A$ and $B$ are the first and second instructions in thread $p_0$, and $C$ and $D$ are the loads of the flag and of the shared variable $z$ in $p_1$, then program order gives rise to $A \hbarrow B$ and $C \hbarrow D$.
We also have that the load of $z$ in $D$ only occurs if the value of the flag observed by thread $p_1$ is \verb|true|, which means it was previously set by thread $p_0$ in $B$.
Therefore, in all runs in which $D$ is observed, $B$ necessarily occurs earlier in the execution.
This necessity does not, however, place $B$ and $D$ in happens-before relation.
Under many flavors of weak memory, the memory accesses between the two threads are not synchronized.
As the example shows, our tentative definition of happens-before as always-occurring-before or necessarily-occurring-before does not work for weak memory systems.  How about for sequential consistent ones?

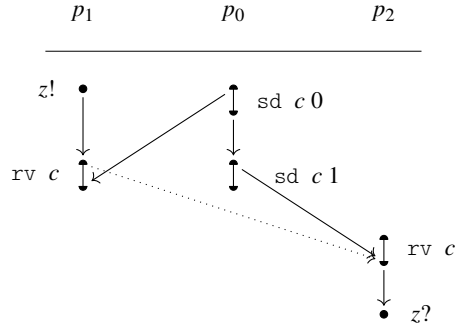
\begin{figure}[htb]
\centering
\input{figures/conditional_race}
\caption{Partial order on conditional-race example.\label{fig:conditional.race.po}}
\end{figure}

In the program of Figure~\ref{fig:conditional.race.prog}, thread $p_0$ sends values $0$ and $1$ into channel $c$ consecutively.
Concurrently, thread $p_1$ writes $42$ to a shared variable $z$ and receives from the channel, while thread $p_2$ first receives from the channel and conditionally reads from $z$.
From this program, we construct an example in which events are necessarily ordered but are not in happens-before---even if we assume sequential consistency.
\begin{figure}[htb]
\begin{align*}
  \qquad\qquad\qquad\qquad\qquad
  &\nslgoroutine{p_0}{\gsend{c}{0};\gsend{c}{1}} \\ 
  &\nslgoroutine{p_1}{\store{z}{42};\ \greceive{c}} \\ 
  &\nslgoroutine{p_2}{\lets r := \greceive{c}\ \ins\ \ifs r=1\ \thens\ \load{z}}
\end{align*}
\caption{Conditional race example.\label{fig:conditional.race.prog}}
\end{figure}
To illustrate this point, let us consider an execution of the program.  Let $\event{p}{o}$ be a trace event capturing the execution of operation $o$ by threads $p$.
Let also $\wrop{z}{}$ and $\rdop{z}{}$ represent a write and read operation
on the shared variable $z$, and $\sdop{}{c}{}$ and $\rvop{}{c}{}$ represent
send and receive operations on channel $c$.  
Assuming channel capacity $\len{c} \geq 2$, the sequence below is a possible trace obtained from the execution of the program.  Note that the if-statement's reduction is interpreted as an internal or silent transition:
\begin{align}
\sdev{p_0}{}{c}{}~~
\sdev{p_0}{}{c}{}~~
\wrev{p_1}{z}{}~~
\rvev{p_1}{}{c}{}~~
\rvev{p_2}{}{c}{}~~
\rdev{p_2}{z}{}
\label{trace:conditional.race}
\end{align}
Given that $p_1$ receives from $c$ before $p_2$ does, the value received by $p_2$ must be $1$ as opposed to $0$.  Therefore, $p_2$ takes the branch and reads from the shared variable $z$.
Figure~\ref{fig:conditional.race.po} shows the partial order on events for this execution.
Program order is captured by the vertical arrows in the diagram; channel communication is captured by the solid diagonal arrows.  
%
%
As per discussion in Section~\ref{sec:race.detector.channels}, we make the distinction between a channel operation and its completion.  
A channel operations is depicted as two half-circles; the operation's completion is captured by the bottom half-circle.  That way, a send (top of the half-circle) happens-before its corresponding receive completes (bottom half).

Now, given that the send operations are in happens-before, meaning $\sdev{p_0}{}{c}{0} \hbarrow \sdev{p_0}{}{c}{1}$, and that channels are First-In-First-Out (FIFO), the reception of value $0$ from $c$ must occur before the reception of $1$.  This requirement is captured by the dotted arrow in the diagram.  However, according to the semantics of channel communication (i.e. rules~(\ref{eq:gomm.hb.chan.forward}) and~(\ref{eq:gomm.hb.chan.backward}) of page~\pageref{eq:gomm.hb.chan.forward}), this order does not impose a happens-before relation between the
receiving events.
In other words, there exist events that are necessarily ordered, but not in happens-before relation to one another.

\medskip

The failure of our tentative definition of happens-before as necessarily-occurring-before, given early in this section, has subtle implications as discussed next.

\subsection{Happens-before, traces, and commutativity of operations}

Traces come from observing the execution of a program and are expressed as strings of events.   In a concurrent system, however, events may not be causally related, which means that the order of some events is not pre-imposed.  In reality, instead of sequences, events in a concurrent system form a partially ordered set (see Figure~\ref{fig:conditional.race.po} for an example).  As advocated by \citet{mazurkiewicz:tracetheory}, it is useful to combine sequential observations with a dependency relation for studying ``the nonsequential behaviour of systems via their sequential observations.''  
By defining an \emph{independence relation} on events, it is possible to derive a notion of equivalence on traces: two traces are equivalent if it is possible to transform one into the other ``by repeatedly commuting adjacent pairs of independent operations''~\cite{katz.peled:defining}.

One way to define independence is as follows:
Given a run $R_i \trans{a} \cdot \trans{b} R$, we say that $a$ and $b$ are independent if $R_i \trans{b} \cdot \trans{a} R$,
meaning,
\begin{itemize}
\item $b$ is enabled at $R_i$,
\item $a$ is enabled at $R_i \trans{b} \cdot$, and
\item there exists an $R'$ such that $R_i \trans{b} R' \trans{a} R$.
\end{itemize}
Clearly, if $a$ happens-before $b$, then $a$ and $b$ cannot be swapped in a trace.  So, independence between two events means (at least) the absence of happens-before relation between them.  But happens-before is not all that needs to be considered in the definition of independence.

When translating a partial order of events to a trace, not every linearization that respects the happens-before relation is a valid trace.  Some linearizations of the partial order may not be ``realizable'' by the operational semantics.  In other words, there can be traces that abide by the happens-before relation but that cannot be generated from the execution of a program.
For example, we can obtain the following linearization given the partial order of Figure~\ref{fig:conditional.race.po}:
\begin{align}
\sdev{p_0}{}{c}{0}~~
\sdev{p_0}{}{c}{1}~~
\rvev{p_2}{}{c}{}~~
\wrev{p_2}{z}{}~~
\rdev{p_1}{z}{}~~
\rvev{p_1}{}{c}{}.
\label{trace:conditional.race.not.realizable}
\end{align}
This linearization respects the partial order based on the happens-before relation: program order is respected, so is the relation between sends and their corresponding receives.  However, this linearization breaks the first-in-first-out assumption on channels.
FIFO is broken because, in order for $p_2$ to read from $z$, it must be that it received the value of $1$ from the channel.  But $p_2$ is the first thread to receive from the channel and, since $0$ was the first value into the channel, it must also have been the first value read from the channel.  Therefore, the linearization in Trace~\ref{trace:conditional.race.not.realizable} is not ``realizable'' by the operational semantics.
While happens-before restricts the commutation of trace operations, there exist other operations that are ordered (though not ordered by happens-before) and that, consequently, must not commute.

The difficulty in conciliating the commutativity of trace events with the happens-before relation remains counterintuitive today, even though its origins are related to an observation made years ago in a seminal paper by \citet{lamport:time}.  In the paper, \citeauthor{lamport:time} points out that ``anomalies'' can arise when there exist orderings that are external to the definition of happens-before---see the ``Anomalous Behavior'' section of~\cite{lamport:time}.  
In order to avoid these anomalies, one suggestion from the paper is to expand the notion of happens-before so that, if $a$ and $b$ are necessarily ordered, then $a$ and $b$ are also in happens-before.

Let us analyze the consequences of rolling FIFO notions into the definition of happens-before.
Given the example of Figure~\ref{fig:conditional.race.prog}, since the sends are ordered in a happens-before relation, and the channel is FIFO, one can argue that the receive events should also be ordered by happens-before.  According to this argument, we ought to promote the dotted line in Figure~\ref{fig:conditional.race.po} to a solid $\hbarrow$ arrow.  This modification would make the example well-synchronized.  
In one hand, given that the write to $z$ by $p_1$ and the read from $z$ by $p_2$ are \emph{always} separated by events (by the two receive events in specific), interpreting the two memory accesses as being synchronized seems rather fitting:  the two memory accesses cannot happen simultaneously, nor can they exist side-by-side in a trace.

There are downsides to this approach.  For one, the resulting semantics deviates from Go's, but, more importantly, such a change does impact synchronization in counter intuitive ways.
Specifically, making the dotted arrow a happens-before arrow would imply that a receiver (in this case $p_2$) can learn about prior events that are not known by the corresponding sender.
If the dotted arrow is promoted to a synchronization arrow, the write $\wrev{p_1}{z}{}$ is communicated to $p_2$ via $p_0$ without $p_0$ itself being ``aware'' of the write.  In other words, the write identifier is transmitted via $p_0$ but is not present in $p_0$'s happens-before set.

We follow Go and allow for some events to always occur in order without affecting synchronization.
Consequently, such ordered events are not considered to be in \emph{happens-before} order.
A less clear consequence, however, is that races can longer be defined as simultaneous (or side-by-side) accesses to a shared variable.  This point is explored next.

\subsection{Manifest data races}
\label{sec:race.discussion.independence.and.manifest.dr}
Section~\ref{sec:race.background} mentioned the concept of manifest data race; below we give a concrete definition.

\begin{definition}[Manifest data race]
  \label{definition:race.manifest.race.configuration}
  \index{data race} \index{data race!manifest} \index{properly manifest} A
  well-formed configuration $R$ contains a \emph{manifest data race} if
  either hold:
  \begin{align*}
  R \trans{\wrev{p_1}{z}{}} \text{~and~} R \trans{\wrev{p_2}{z}{}}
    & 
    & \text{(manifest write-write race on $z$)} \\
  R \trans{\rdev{p_1}{z}{}} \text{~and~} R \trans{\wrev{p_2}{z}{}} &
    & \text{(manifest read-write race on $z$)}
  \end{align*}
  for some $p_1\not=p_2$.
\end{definition}

Manifest data races can also be defined on traces.

\begin{definition}[Manifest data race]
  \label{definition:race.manifest.race.trace}
  \index{data race} \index{data race!manifest} \index{data
    race!manifest!trace}\index{properly manifest} A well-formed trace $h$
  contains a \emph{manifest data race} if either
  \begin{align*}
  \wrev{p_1}{z}{}~ &\wrev{p_2}{z}{}\,    &  \text{(manifest write-after-write)} \\
  \wrev{p_1}{z}{}~ &\rdev{p_2}{z}{}\,   &  \text{(manifest read-after-write)} \\
  \rdev{p_1}{z}{}~ &\wrev{p_2}{z}{}    &  \text{(manifest write-after-read)}
  \end{align*}
  are a sub-sequence of $h$ and where $p_1\not=p_2$).
\end{definition}

While manifest races are obvious, races in general may involve accesses that are arbitrarily ``far apart'' in a linear execution.
By bring conflicting accesses side-by-side, we could show irrefutable evidence of a race that, otherwise, may be obscured in a trace.
Let $h \tracepreo h'$ represent the fact that $h'$ is derivable from $h$ by the repeated commutation of adjacent pairs of independent operations.  If $h \tracepreo h'$ and $h'$ contains a manifest data race, then we say $h$ contains a data-race.  This definition of races seems unequivocal.  From here, soundness and completeness of a race detector may be defined as such:

\begin{theorem}{(Soundness)}
\label{lemma:race.soundness}
If $S_0 \trans{h}$ is a run flagged by a data-race detector, then $h \tracepreo h_{dr}$ with $h_{dr}$ containing a manifest data-race.
\end{theorem}

\begin{theorem}{(Completeness)}
\label{lemma:race.completeness}
Let $S_0 \trans{h}$ be a run such that $h \tracepreo h_{dr}$ and $h_{dr}$ contains a manifest race.  Then $S_0 \trans{h}$ is flagged by the data-race detector.
\end{theorem}

Theorems~\ref{lemma:race.soundness} and~\ref{lemma:race.completeness} are also clear and unequivocal.  More importantly, they link two world views: the view of races as unsynchronized accesses with respect to the happens-before relation and a view of races in terms of commutativity of trace events \textit{\`a la} Mazurkiewicz.
The problem with the concept of manifest data race and Theorems~\ref{lemma:race.soundness} and~\ref{lemma:race.completeness}, however, is that when the definition of independence is made to respect FIFO order as well as the happens-before relation, the notion of manifest data race is no longer attainable.  In other words, given a definition of independence which respects FIFO and happens-before, there exist racy traces from which a manifest data race is not derivable.

The program of Figure~\ref{fig:conditional.race.prog} gives rise to such an example.  The access to $z$ by $p_2$ only occurs if $p_2$ receives the second message sent on the channel.  In other words, the existence of event $\rdev{p_2}{z}{}$ in a trace is predicated on the order of execution of channel operations: $p_2$ only reads from $z$ if the other thread, $p_1$, receives from $c$ before $p_2$ does.\footnote{In this example, we use the value of the message received on a channel to branch upon.  But since a receive from a channel changes a thread's ``visibility'' of what is in memory, it is possible to craft a similar example in which all message values are \texttt{unit} but in which a thread's behavior changes due to a change in the ordering of the receives.}  This requirement places the receive operations between the memory operations.  Therefore, a trace in which $\wrev{p_1}{z}{}$ and $\rdev{p_2}{z}{}$ are side-by-side is not attainable.  Yet, as discussed previously, the accesses to $z$ are not ordered by happens-before, and, therefore, are concurrent.  Since the accesses are also conflicting, they constitute a data race.

It seems that Mazurkiewicz traces are ``more compatible'' with confluence checking than data-race checking.  In data-race checking, there are non-confluent runs that do not exhibit data races; these runs are non-confluent because they have ``races on channels.''  In our example, the two receives from $p_1$ and $p_2$ are in competition for access to the channel.  These receive operations are concurrent and non-confluent.
Finally, the example also hints at the perhaps more fundamental observation: that races have little to do with \emph{simultaneous} accesses to a shared variable but instead with \emph{unsynchronized} accesses.  While simultaneous accesses are clearly unsynchronized, not all unsynchronized accesses may be made simultaneous.\footnote{There may not exist a configuration from which two transitions are possible; transitions that involve conflicting memory accesses.  Yet, it is possible for two access separated ``in time'' to be unsynchronized.}

%% file: figures/conditional_race.tex
\begin{tikzpicture}[]
    \node (p0) at (0,0)  {$p_0$};
    \node (p1) at (-2,0) {$p_1$};
    \node (p2) at (2,0)  {$p_2$};
    \draw[line width=0.1mm] (-2.5,-0.5) -- (2.5,-0.5);

    \dop{p0sd0}{$\sdop{}{c}{0}$}{0}{-1}{right}
    \dop{p0sd1}{$~~\sdop{}{c}{1}$}{0}{-2}{right}
    \pop{p1z}{$\wrop{z}{}$}{-2}{-1}{left}
    \dop{p1rv}{$\rvop{}{c}{}$}{-2}{-2}{left}
    \dop{p2rv}{$\rvop{}{c}{}$}{2}{-3}{right}
    \pop{p2z}{$\rdop{z}{}$}{2}{-4}{right}

    \draw [->] (p0sd0e) -- (p0sd1s) node [near end,right]  {$\myhb$};
    \draw [->] (p0sd0s) -- (p1rve)  node [near end,right]  {$\myhb$};

    \draw [->] (p1z)   -- (p1rvs)  node [near end,left]  {$\myhb$};

    \draw [->] (p2rve)  -- (p2z)   node [near end,right] {$\myhb$};
    \draw [->] (p0sd1s) -- (p2rve)  node [near end,right]  {$\myhb$};

    \draw [->,dotted] (p1rvs) -- (p2rve)  node [near end,left]  {};
\end{tikzpicture}

%% file: related.tex
\section{Related work}
\label{sec:race.related}

Race detection via the analysis of source code is an undecidable problem.
Regardless, race detectors via the static analysis of source code \cite{naik.aiken.whaley:effective, voung.jhala.lerner:relay, blackshear2018racerd} exist and have found application in industry.  More recently, \citet{blackshear2018racerd} implement a static analysis tool called \racerd\ to help the parallelization of previously sequential Java source code.
The tool over approximates the behavior of programs and can, thereby, reject programs that turn out to be data-race free.  This over approximation was not a hindrance, as even conservative parallelization efforts can lead to gains over purely sequential code.

By and large, however, instead of flagging races in a program as a whole, race detectors have resorted the analysis of particular \emph{runs} of a program.  To that end, detectors instrument the program so that races are either flagged during execution, in what is called \emph{on-line} or \emph{on-the-fly} race detection, or on logs captured during execution and analyzed \textit{postmortem}.
Even still, dynamic race detection is NP-hard~\cite{netzer1990complexity} and many techniques have been proposed for detection at scale.
Broadly, these techniques involve static analysis used to reduce the number of runtime checks~\cite{flanagan.freund:fasttrack}\cite{rhodes2017bigfoot}, and heuristics that trade false-positive~\cite{savage*:eraser, pratikakis.foster.hicks:locksmith, choi*:efficient} or false-negative rates~\cite{marino2009literace} for better space/time utilization.  
For example, by allowing races to sometimes go undetected, \emph{sampling} race detectors let go of completeness in favor of lower overheads. One common heuristic, called the \emph{cold region hypothesis}, is to sample more frequently from less executed regions of the program.  This rule-of-thumb hinges on the assumption that faults are more likely to already have been identified and fixed if they occur in the hot regions of a program~\cite{marino2009literace}.
Alternatively, by going after a proxy instead of an actual race, imprecise race detectors let go of soundness.  The prominent examples here are Eraser's LockSet~\cite{savage*:eraser} and Locksmith~\cite{pratikakis.foster.hicks:locksmith}, which enforce a lock-based synchronization discipline.  A violation of the discipline is a \emph{code smell} but not necessarily a race.
The amalgamation of different approaches have also been investigated, leading to \emph{hybrid race detectors}. 
For example, \citet{ocallahan2003hybrid} combined LockSet-based detection with happens-before information reconstructed from vector clocks; \citet{choi*:efficient} extended LockSet to incorporate static analyses.

Another avenue of inquiry has lead to \emph{predictive race detection}~\cite{smaragdakis2012sound, huang2014maximal}, which attempts to achieve higher detection capabilities by extrapolating beyond individual runs.
\citet{huang2014maximal} incorporate abstracted control flow information and formulate race detection as a constraint solving problem.
With the goal of observing more races per run, \citet{smaragdakis2012sound} introduce a new relation, called \emph{causally-precedes}, which is a generalization of the happens-before relation.

A number of papers address race detection in the context of channel
communication~\cite{cypher1995efficient, damodaran1993nondeterminancy, terauchi2008capability}.
Some of the papers, however, do not speak of shared memory but, instead, define
races as conflicting channel accesses.  In that setting, the lack of
conflicting accesses to channels imply determinacy.
A different angle is taken by \citet{terauchi2008capability}, who,
among different kinds of channels, define a buffered channel whose buffer
is overwritten by every write (i.e. send) but never modified by a read
(i.e. receive).  This kind of channel, referred to as a \emph{cell}, behaves,
in essence, as shared memory.
The goal of \citet{terauchi2008capability} is, still, determinacy.
Having conflated the concept of shared memory as a channel, determinacy is
then achieved by ensuring the absence of conflicting accesses to channels.
Our goal, however, is different: we aim to detect data-races but do not 
want to go as far as ensuring determinacy.  Therefore, our approach allows
``races'' on channel accesses.
From a different perspective, however, the work of \citet{terauchi2008capability} can be seen as complementary to ours:  We conjecture that their type system can serve as the basis for a static data-race detector.

Among the dynamic data-race detection tools from industry, \citet{banerjee2006theory} discuss different race detection algorithms including one used by the Intel Thread Checker.  The authors describe \emph{adjacent conflicts}, which is similar to our notion of side-by-side or manifest data race.  The paper also classifies races similar to our WaR, RaW, and WaW classification.

Go has a race detector integrated to its tool chain
\cite{golang.racedetector}.  The \verb!-race! command-line flag instructs the
Go compiler to instrument memory accesses and synchronization events.  The
race detector is built on top of Google's sanitizer
project~\cite{google.sanitizer} and TSan in particular
\cite{serebryany2009threadsanitizer, google.threadsanitizer}.
TSan is part of the LLVM's runtime libraries
\cite{serebryany2011dynamic, llvm.threadsanitizer};
it works by instrumenting memory accesses and monitoring locks
acquisition and release as well as thread forks and joins.
Note, however, that channel communication is the vehicle for achieving
synchronization in Go.  Even though locks exist, they are part of a package,
while channels are built into language.  Yet, the race detector for Go sits at
a layer underneath.  In this paper we study race detection with channel communication taking a central role.
Also, different from TSan, we employ propose a technique based on what we call
\emph{happens-before sets} as opposed to vector clocks.  The consequences of
this decision is discussed in detail on Section~\ref{sec:race.comparison}.

It is also relevant to point out that, in the absence of the DRF-SC guarantee, one may resort to finding data races involving weak memory behavior.
Since the full \CCpluspluseleven\ memory model can harbor such races, and with the goal of finding data races in production level code, \citet{lidbury2017dynamic} extend the ThreadSanitizer (TSan)~\cite{serebryany2009threadsanitizer, google.threadsanitizer} to support a class of non-sequentially consistent executions.

%% file: conclusion.tex
\section{Conclusion}
\label{sec:race.conclusion}

We presented a dynamic data-race detector for a language in the style of Go: featuring channel communication as sole synchronization primitive.
The proposed detector records and analyzes information locally and is well-suited for online detection.  
\iftrace
The operational semantics of the detector is given along with a proof of soundness and completeness.  
The proofs relate reductions in the semantics to events in a trace grammar.
A trace $h$ contains a data race if $h'$ contains a \emph{manifest data race} (see Definition~\ref{definition:race.trace.manifest.race}) and $h'$ can be derived from $h$.
Thus, the concept of manifest data race and the derive-from relation $\tracepreo$ give us an idealized race detector that serves as specification against which we establish the correctness of proposed race detector's operational semantics.
Soundness is proven by showing that, if the race detector flags an execution as racy, there exists a manifest data-race in the trace underlying the given program's execution.
Similarly, for completeness, if a trace containing a manifest data-race can be derived from an execution, then the execution is flagged as racy by the detector.
\fi

Our race detector is built upon a previous result~\cite{fava*:oswmm-chan-journal}, where we formalize a weak memory model inspired by the Go specification~\cite{www:gomemorymodel}.  In that setting, we recorded memory read- and write-events that were in happens-before relation with respect to a thread's present operation.  This information was stored in a set called $\idshb$ or the happens-before set of a thread, and it was used to regulate a thread's visibility of memory events.  The core of the paper was a proof of the DRF-SC guarantee, meaning, we proved that the proposed relaxed memory model behaves sequentially consistently in the absence of data races.  The proof hinges on the fact that, in the absence of races, all threads agree on the contents of memory; see the \emph{consensus lemma} in~\cite{fava*:oswmm-chan-journal}.  The scaffolding used in the proof of the consensus lemma contains the ingredients used of the race detectors presented in this paper.
Based on our experience, we conjecture that one may automatically derive a race detector given a weak memory model and its corresponding proof of the DRF-SC guarantee.

In the DRF-SC the proof of~\cite{fava*:oswmm-chan-journal}, we show that if a program is racy, it behaves sequentially consistent up to the point in which the first data-race is encountered.
In other words, this first point of divergence sets in motion all behavior that is not sequentially consistent and which arise from the weakness in the memory model.
With this observation, we argue that a race detector can operate under the assumption of sequential consistency.  This is a useful simplification, as sequential consistent memory is conceptually much simpler than relaxed memories.
If the data-race detector flags the first evidence of a data-race, then program behavior is sequentially consistent up to that point.

\iftrace
\ifworking
{\color{blue}
We see the formalized race detector as:
1) A documentation for what it means to be a dynamic race-detector for a language featuring channel communication as sole synchronization primitive.
2) A specification for a sound and complete race detector for a Go-like language.
Go already has a dynamic race-detector which has undergone many years of engineering effort both inside and outside of Google~\cite{golang.racedetector}.  The Go race detector is built on top of Google's sanitizer project~\cite{google.sanitizer} and the thread sanitizer TSan in particular~\cite{serebryany2009threadsanitizer}.
We see our proposed race detector as a specification against which the Go race-detector's implementation can be compared and contrasted.
It may be worth while, provided sufficient resources, to attempt to bridge the gap between the specification and the implementation of the two race detectors.
}
\fi
\fi

Avenues for future work abound.  
In contrast to data-race detectors based on vector clocks, our approach using happens-before sets does not provide as terse of a representation for the collection of memory events performed by a thread in between synchronization points.  In effect, out approach has a larger foot-print, which ought to be mitigated.  On the other hand, our thorough expunging of stale information can serve as inspiration to vector clock based approaches, which allow for the accumulation of stale information---see Section~\ref{sec:race.discussion}.
Another extension would be to statically analyze a target program with the goal of removing dynamic checks or ameliorating the detector's memory consumption.  Here, we may be able to borrow from the research on static analysis for dynamic race-detection in the context of lock-based synchronization disciplines.

\iftrace
\ifworking
{\color{blue}
The formalization of the independence relation can serve as a stepping stone into model checking.  ``Verification can benefit from the fact that the effect of two independent concurrent operations is typically commutative.  This commutativity among operations allows reducing the number of program states explicitly considered''~\cite{katz.peled:defining}.
}
\fi
\fi

%% file: strongsemantics.tex
\section{Strong semantics}
\label{sec:gomm.race.semantics.strong}

\ifok
For completeness sake and for reference, we include here the operational
semantics \emph{without} augmenting it with any information relevant for
race checking. It is thereby a conventional operational semantics and
corresponds to the strong semantics from~\cite{fava*:oswmm-chan-journal}.

\renewcommand{\ewritesl}[3]                       {\bananas{#2{:=}#3}}                      
\renewcommand{\ngoroutine}[3]                     {\angles{#3}}      

\begin{figure}[ht]
  \centering
  \input{rules/sugar}
  \caption{Syntactic sugar\label{fig:race.syn.sugar}}
\end{figure}

The surface syntax is unchanged from Figure~\ref{fig:race.grammar}. The
operational semantics is formulated using run-time configurations as given
in equation~(\ref{eq:gomm.race.naked.configs}). 

\input{./rules/datarace/configs-raw-waw}

For race detection, we used the ``same'' run-time syntax, except that they
were augmented with additional information (cf.\ equation for the
intermediate formulation~(\ref{eq:gomm.race.configs-simple}) of the race
detecting semantics resp.\ equation~(\ref{eq:gomm.race.gc.configs}). Compared to the race detecting
semantics, the configurations carry less information. In particular, the
recorded events don't carry identifying labels and threads don't keep track
of happens-before information as for the race checker.\footnote{Note in
  passing, also in the formalization of the weak semantics in~\cite{fava*:oswmm-chan-journal}, the threads keep track of happens-before
  information. Here, the additional information is needed to do race
  detection on the strong semantics, where the semantics itself works
  without that information, whereas in~\cite{fava*:oswmm-chan-journal}, the
  additional information is required to describe the (weak) semantics
  itself.}

\subsection{Structural congruence}
\label{sec:gomm.race.cong}

\index{$\congstruct$ (structural congruence)}
\index{structural congruence}

Configurations are interpreted up-to structural congruence, only: Parallel
composition is associative and commutative, with the empty configuration as
neutral element. The $\nu$-binder is used to manage the scopes for
dynamically created names. Besides that, syntax is considered tacitly up-to
renaming of bound names, in particular, $\nu$-bound names. 

\begin{table}[ht]
  \centering
  \begin{displaymath}
    \input{rules/congstruct}
  \end{displaymath}
  \caption{Structural congruence\label{tab:gomm.race.congstruct}}
\end{table}

Dynamically created names are channel names. In the augmented semantics,
where processes are named and also events carry a label, also names for
those entities can be created on-the-fly and they are subject to the
congruence rules for $\nu$-bound names.

\subsection{Local steps}
\label{sec:gomm.race.steps.local}
\index{local step}
The rules from Figure~\ref{fig:race.steps.local} concern reduction steps
that don't affect the memory or involve channel communication.

\begin{figure}[!ht]
  \centering
  \begin{ruleset}
    \input{rules/datarace/steps-hb-local}
  \end{ruleset}
  \caption{Local steps\label{fig:race.steps.local}}
\end{figure}

\fi 

\subsection{Memory interactions and channel communication}
\label{sec:gomm.race.steps.nonlocal}

Reading and writing, the two basic memory interactions, are covered in
Figure~\ref{fig:race.steps.smem.naked} and channel communication in Figure~\ref{fig:race.steps.chans.naked}. Compared to the semantics for race
detection (cf.\ Figures~\ref{fig:race.steps.smem.full} and~\ref{fig:race.steps.chans.full}), the semantics here is done without extra
information and book-keeping of happens-before information. Related to
that, the recorded events don't carry any names to identify the event.

\begin{figure}[!ht]
  \centering
  \begin{ruleset}
    \input{rules/strong/steps-smem}
  \end{ruleset}
  \caption{Read and write steps\label{fig:race.steps.smem.naked}}
\end{figure}

For the channel communication in Figure~\ref{fig:race.steps.chans.naked},
no more happens-before information is communicated. Especially the forward
channel carries only the communicated value $v$ (cf.\ rule \rn{R-Send}
and \rn{R-Rec}). To realize the boundedness of the channels, the semantics
still maintains the two parts of a channel: the forward channel for
communication, and the backward channel for ``flow-control.''  The backward
channel $c_b$ does not carry any information, just the number of entries
representing still empty slots in the forward channel. We use the unit
value $()$ for that, and initially, the backward channel is filled with a
number of $()$'s corresponding to the capacity of the channel (see rule
\rn{R-Make}).

For channel communication, the semantics distinguished between synchronous
communication, i.e., a ``rendezvous'' over a channel of capacity 0, and
asynchronous communication, with a channel of non-zero, but finite
capacity. For the asynchronous case, both sending and receiving

\begin{figure}[!ht]
  \renewcommand{\lstateempty}{()}
  \renewcommand{\esend}[3]{#2}
  \centering
  \begin{ruleset}
    \input{rules/strong/steps-chans}
  \end{ruleset}
  \caption{Channel communication\label{fig:race.steps.chans.naked}}
\end{figure}


%% file: rules/sugar.tex
\begin{displaymath}
  \begin{array}[t]{rcl@{\quad}l}
    e ;~ t & \eqdef & \lets r = e \ins t & \text{when $r \notin \fv{t})$}
    \\
    \stops & \eqdef & \selects_0 
  \end{array}
\end{displaymath}


%% file: rules/datarace/configs-raw-waw.tex
\begin{equation}
  \label{eq:gomm.race\VERSION.configs}
  \begin{array}[t]{l}
  R \bnfdef \ngoroutine{p}{\idshb}{t}
  \bnfbar
  \ewritesl{m}{z}{v}
  \bnfbar 
  \emptyconf 
  \bnfbar
  R \parallel R\
  \bnfbar 
  \gchan{c}{q}
  \bnfbar 
    \nu n\  R\  .
  \end{array}
\end{equation}
%

%% file: rules/congstruct.tex
\begin{array}[t]{rcll}
  R_1 \parallel R_2 & \congstruct & R_2 \parallel R_1 
  \\
  (R_1 \parallel R_2) \parallel R_3 & \congstruct & R_1 \parallel (R_2 \parallel R_3)
  \\
  \emptyconf \parallel R & \congstruct & R
  \\
  R_1 \parallel \nuin{n}{R_2} & \congstruct &  \nuin{n}{(R_1 \parallel R_2)} & \text{\qquad if $n \notin\fn(R_1)$}
  \\
  \nuin{n_1}{\nuin{n_2}{R}}  & \congstruct &  \nuin{n_2}{\nuin{n_1}{R}}
\end{array}

%% file: rules/datarace/steps-hb-local.tex
\infax{R-Red}{
  \lets x = v \ins t \transloc{} t\substfor{v}{x}
}

\\

\infax{R-Let}{
  \lets x_1 = (\lets x_2 = e \ins t_1)\ins t_2 \transloc{} \lets x_2 = e \ins (\lets x_1 = t_1 \ins t_2)
}

\\

\infax{R-Cond$_1$}{
  \ifs\ \trues\  \thens t_1 \elses t_2
  \transloc{}
  t_1
}

\infax{R-Cond$_2$}{
  \ifs\ \falses\  \thens t_1 \elses t_2
  \transloc{}
  t_2
}

%
%

%% file: rules/strong/steps-smem.tex
\infax{R-Write}{
  \ngoroutine{p}{\idshb}{\store{z}{v'}; t}
  \parallel
  \ewritesl{m}{z}{v}
  \trans{}
  \ngoroutine{p}{\idshb'}{t}
  \parallel
  \ewritesl{m'}{z}{v'}
}

\ruleskip

\infax{R-Read}{
  \ngoroutine{p}{\idshb}{\lets r=\ \load{z} \ins t}
  \parallel
  \ewritesl{m}{z}{v}
  \trans{}
  \ngoroutine{p}{\idshb}{\lets r=\ v \ins t}
  \parallel
  \ewritesl{m}{z}{v}
}


%% file: rules/strong/steps-chans.tex
\infrule{R-Make}{
  q = [\lstateempty,\ldots,\lstateempty]
  \andalso
  \sizeof{q} = v
  \andalso
  \fresh(c)
}{
  \ngoroutine{p}{\idshb}{\lets r = \ \makechans{\T}{v} \ins t} 
  \widetrans{}
  \nuin{c}{(  \ngoroutine{p}{\idshb}{\lets r = c \ins t}   
  \parallel
  \gchanf{c}{}
  \parallel
  \gchanb{c}{q})
  }
}

\ruleskip

\infrule{R-Send}{
  \lnot \gisclosed (\gchanf{c}{q_2})
}{
    \gchanb{c}{q_1::()} \parallel  
    \ngoroutine{p}{\idshb}{\gsend{c}{v};t} \parallel  
    \gchanf{c}{q_2}
    \widetrans{}
    \gchanb{c}{q_1} \parallel  
    \ngoroutine{p}{\idshb'}{t}
    \parallel \gchanf{c}{\esend{}{v}{\idshb}::q_2}
}

\ruleskip

\infrule{R-Rec}{
  v \not= \eot
}{
  \begin{array}{rcl}
    \gchanb{c}{q_1} \parallel &
    \ngoroutine{p}{\idshb}{\lets r = \greceive{c}\ins t} &
    \parallel \gchanf{c}{q_2::\esend{}{v}{\idshb''}} \widetrans{} \\
    \gchanb{c}{\idshb::q_1} \parallel &
    \ngoroutine{p}{\idshb'}{\lets r = v \ins t} &
    \parallel \gchanf{c}{q_2}
  \end{array}
}

\ruleskip

\infrule{R-Rec$_\eot$}{
}{
  \ngoroutine{}{}{\lets r = \greceive{c}\ins t}
  \parallel
  \gchanf{c}{\eot}
  \widetrans{}
  \ngoroutine{p}{}{\lets r = \eot \ins t}
  \parallel
  \gchanf{c}{\eot}
}

\ruleskip

\infrule{R-Rend}{
}{
  \begin{array}{rcll}
    \gchanb{c}{} \parallel &
    \ngoroutine{p_1}{}{\gsend{c}{v}; t} &
    \parallel \ngoroutine{}{p_2}{\lets r = \greceive{c}\ins t_2} &
    \parallel \gchanf{c}{} \widetrans{} \\
    
    \gchanb{c}{} \parallel &
    \ngoroutine{p_1}{}{t} &
    \parallel \ngoroutine{p_2}{}{\lets r = v\ins t_2} &
    \parallel \gchanf{c}{}
  \end{array}
}

\ruleskip

\infrule{R-Close}{
  \lnot \gisclosed ( \gchanf{c}{q})
}{
  \ngoroutine{p}{}{\gcloses(c); t}
  \parallel
  \gchanf{c}{q} 
  \widetrans{}
  \ngoroutine{}{p}{t}
  \parallel
  \gchanf{c}{\eot:: q}
}
%

%% file: append.tex
\section{Rules for the select statement}
\label{sec:race.select}

Rules dealing with the select statement semantics are given on
Figure~\ref{fig:race.steps.global.select}.  The \rn{R-Sel-Send} and
\rn{R-Sel-Rec} rules apply to asynchronous channels and are analogous to
\rn{R-Send} and \rn{R-Rec}.  The \rn{R-Sel-Sync} rules apply to open
synchronous channels (i.e. the forward and backward queues are empty).  The
\rn{R-Sel-Rec$\eot$} is analogous to \rn{R-Rec$\eot$}.  Finally, the
default rule (\rn{R-Sel-Def}) applies when no other select rule applies.
\ifworking
\begin{edfnote}{ugly notation}
  The $'$ and $''$ in $\idshb$ of the rules below are a bit
  ugly/inconsistent. Sometimes $'$ is used to the updated one, sometimes
  $''$ or even no prime is used to the old one.
\end{edfnote}
\fi

\begin{figure}[!ht]
  \centering
  \begin{ruleset}
    \input{rules/datarace/steps-hb-select}
  \end{ruleset}
  \caption{Operational semantics:
   Select statement\label{fig:race.steps.global.select}}
\end{figure}


%% file: rules/datarace/steps-hb-select.tex
\infrule{R-Sel-Send}{
  g_i = \gsend{c}{v}
  \andalso
  \lnot \gisclosed (\gchanf{c}{q_f})
  \andalso
  \idshb' = \idshb + \idshb''
}{
  \begin{array}{rcl}
    \gchanb{c}{q_b::\esendack{}{\idshb''}} \parallel  &
    \ngoroutine{p}{\idshb}{\selects_i \lets r_i = g_i \ins t_i} &
    \parallel \gchanf{c}{q_f} \widetrans{} \\

    \gchanb{c}{q_b} \parallel &
    \ngoroutine{p}{\idshb'}{t_i \substfor{\unit}{r_i}} &
    \parallel \gchanf{c}{\esend{}{v}{\idshb})::q_f}
  \end{array}
}

\ruleskip

\infrule{R-Sel-Rec}{
  g_i = \greceive{c}
  \andalso
  q_f = q_f'::\esend{}{v}{\idshb''}
  \andalso
  v \not= \eot
  \andalso
  q_b' = \esendack{}{\idshb}::q_b
  \andalso 
  \idshb' = \idshb + \idshb''
}{
  \begin{array}{rcl}
    \gchanb{c}{q_b} \parallel &
    \ngoroutine{p}{\idshb}{\selects_i \lets r_i = g_i \ins t_i} &
    \parallel \gchanf{c}{q_f} \widetrans{} \\

    \gchanb{c}{q_b'} \parallel &
    \ngoroutine{p}{\idshb'}{\lets r_i = v \ins t_i} &
    \parallel \gchanf{c}{q_f'}
  \end{array}
}

\ruleskip

\infrule{R-Sel-Sync$_1$}{
  g_i = \gsend{c}{v}
  \andalso
  \idshb = \idshb' + \idshb''
  \andalso
  \gchanb{c}{}
  \andalso
  \gchanf{c}{}
}{
  \begin{array}{rl}
    \ngoroutine{p_1}{\idshb'}{\selects_i r_i = g_i \ins t_i}
    \parallel &
    \ngoroutine{p_2}{\idshb''}{\lets r = \greceive{c}\ins t_2}
    \widetrans{} \\

    \ngoroutine{p_1}{\idshb}{t_i \substfor{\unit}{r_i}}
    \parallel &
    \ngoroutine{p_2}{\idshb}{\lets r = v\ins t_2}
  \end{array}
}

\ruleskip

\infrule{R-Sel-Sync$_2$}{
  g_i = \greceive{c}
  \andalso
  \idshb = \idshb' + \idshb''
  \andalso
  \gchanb{c}{}
  \andalso
  \gchanf{c}{}
}{
  \begin{array}{rl}
    \ngoroutine{p_1}{\idshb'}{\gsend{c}{v};t_1} \parallel &
    \ngoroutine{p_2}{\idshb''}{\selects_i \lets r_i = g_i \ins t_i}
    \widetrans{} \\

    \ngoroutine{p_1}{\idshb}{t_1} \parallel &
    \ngoroutine{p_2}{\idshb}{\lets r_i = v\ins t_i}
  \end{array}
}

\ruleskip

\infrule{R-Sel-Sync$_3$}{
  g_i = \gsend{c}{v}
  \andalso
  g_j = \greceive{c}
  \andalso
  \idshb = \idshb' + \idshb''
  \andalso
  \gchanb{c}{}
  \andalso
  \gchanf{c}{}
}{
  \begin{array}{rl}
    \ngoroutine{p_1}{\idshb'}{\selects_i \lets r_i = g_i \ins t_i}
    \parallel &
    \ngoroutine{p_2}{\idshb''}{\selects_j \lets r_j = g_j \ins t_j}
    \widetrans{} \\

    \ngoroutine{p_1}{\idshb}{t_i \substfor{\unit}{r_i}}
    \parallel &
    \ngoroutine{p_2}{\idshb}{\lets r_j = v \ins t_j}
  \end{array}
}

\ruleskip

\infrule{R-Sel-Rec$_\eot$}{
  g_i = \greceive{c}
  \andalso
  \gchanf{c}{\esend{}{\eot}{\idshb''}}
  \andalso 
  \idshb' = \idshb + \idshb''
}{
  \ngoroutine{p}{\idshb}{\selects_i \lets r_i = g_i \ins t_i}
  \widetrans{}
  \ngoroutine{p}{\idshb'}{\lets r_i = \eot \ins t_i}
}

\ruleskip

\infrule{R-Sel-Def}{
  g_i = \defaults
  \andalso
  \lnot\exists j.~~ i \neq j.~~  \ngoroutine{p}{\idshb}{\selects_j \lets r_j = g_j \ins t_j} \parallel P
    \trans{}
    \ngoroutine{p}{\idshb'}{t'} \parallel P'
}{
  \ngoroutine{p}{\idshb}{\selects_i \lets r_i = g_i \ins t_i}
  \parallel P
  \widetrans{}
  \ngoroutine{p}{\idshb}{t_i \substfor{\unit}{r_i}}
  \parallel P
}
